\documentclass{JHEP3}
\usepackage{amssymb,amsfonts,amsmath,amsthm,latexsym,graphicx,verbatim,epsfig}
\newcommand{\be}{\begin{equation}}
\newcommand{\ee}{\end{equation}}
\newcommand{\bea}{\begin{eqnarray}}
\newcommand{\eea}{\end{eqnarray}}
\newcommand{\la}{\lambda}
\newcommand{\dla}{\delta\lambda}
\newcommand{\drho}{\delta\rho}
\newcommand{\al}{\alpha}
\newcommand{\bt}{\beta}
\newcommand{\sB}{\sqrt{-\Box}}
\newcommand{\om}{\omega}
\newcommand{\pd}{\partial}
\newcommand{\iii}{\int\limits_{-\infty}^{\infty}}
\newcommand{\hA}{\hat{A}}
\newcommand{\hB}{\hat{B}}
\newcommand{\bB}{\bar{B}}

\title{\center{On rolling, tunneling and decaying in some large N vector models}}

\author{
    Vadim Asnin\footnotemark[1], Eliezer Rabinovici\footnotemark[2], Michael Smolkin\footnotemark[3]\\
    Racah Institute of Physics\\
    Hebrew University \\
    Jerusalem 91904,
    Israel\\

    \footnotemark[1] {\tt vadim.asnin@mail.huji.ac.il\\}
    \footnotemark[2] {\tt eliezer@vms.huji.ac.il\\}
    \footnotemark[3] {\tt smolkinm@phys.huji.ac.il}
        }

\abstract{Various aspects of time-dependent processes are studied
within the large N approximation of O(N) vector models in three
dimensions. These include the rolling of fields, the tunneling and
decay of vacua. We present an exact solution for the quantum
conformal case and find a solution for more general potentials when
the total change of the value of the field is small. Characteristic
times are found to be shorter when the time dependence of the field
is taken into account in constructing the exact large N effective
potentials. We show that the different approximations yield the same
answers in the regions of the overlap of the validity. A numerical
solution of this potential reveals a tunneling in which the bubble
that separates the true vacuum from the false one is thick.}

\begin{document}

\section{Introduction}

Evolutionary processes were studied in particle physics during the
whole time of its own evolution. However, in recent years the study
of time-dependent processes has played an increasing role in
cosmology as well as Quantum Field Theory. The spot light has
shifted away from the study of vacua and perturbations around it.
Metastable vacua and their properties are investigated in detail. In
many cases an understanding of the time-dependent processes is
important. In addition to the interesting theoretical issues
involved, the problem emerged in cosmology in general, and in the
cosmology which deals with a notion of a landscape of vacua in
particular. The time-dependent processes determine both the
tunneling to a different potential well and the rolling of a field
down to the region of a minimum of a potential. Similar physical
processes appear also in some field theory models of possible
phenomenological interest, like for instance in the supersymmetric
models with long-living metastable vacua in which SUSY is broken
\cite{Intriligator:2006dd} . Also in the Standard Model and beyond
it such a tunneling followed by a rolling towards the minimum can
occur.

In most cases a detailed study of these processes is difficult.
Systems accounting for gravity and ignoring it were studied. Various
approximations have been used, such as the semiclassical and the
thin wall approximations \cite{Coleman:1977py, Coleman:1980aw,
Banks:2002nm} .

Understanding the exact features of such time-dependent processes is
important for theory and may eventually have practical uses.

For this purpose we conduct a study  in the case of large $N$ vector
models in three space-time dimensions.

Traditionally, quantum field theories with $N$ dynamical variables,
$N\to\infty$, have served mostly as study grounds for extending
one's intuition in handling basic problems of quantum field theories
\cite{Moshe:2003xn}. Approximate results obtained at large $N$
possess many of the properties believed to be true in an exact
solution. Such processes have been studied in the large $N$
approximation of the vector model in $d=4$. Here we obtain new
results and highlight additional aspects for $d=3$, which includes a
quantum conformal case.

In this paper  we investigate various time-dependent processes in
this model. In section \ref{useful_formulas} we review the rich
phase structure of the three dimensional vector models and the
methods of solving them. We next calculate in section
\ref{NonConstantFields} an exact large $N$ effective action which
allows treating time-dependent solutions. In the derivation we take
into account time variations also in the potential terms. We derive
the general equations which govern an evolution of the system and
solve them in two special cases. In the first case (section
\ref{ConformalCase}) we consider a rolling of the field from the top
of a potential $\phi^6$. The corresponding theory is also quantum
conformal \cite{Bardeen:1983rv}, and it makes a solution accessible.
We show that the field continues to escape to infinity in a finite
time. A time is shorter, however, than that obtained in an
approximation where the time dependent effects in the potential are
not accounted for. This situation is of interest in studying
possible holographic descriptions of Big Crunch singularities
\cite{Hertog:2004rz} .

In the second case (section \ref{SmallChangesInFields}) we consider
a rolling in a general potential in an approximation of small
changes in fields along the evolution. We derive a characteristic
time of run away from a potential maximum and a frequency of small
oscillations around a minimum. The frequency of oscillations is
bigger than the corresponding frequency obtained for an effective
potential derived assuming static solutions only.

We also study a possibility of a tunneling in the system (section
\ref{TunnelingSection}). We show that a solution in this case looks
like an expanding bubble of a complicated shape different than that
resulting in  the thin-wall approximation \cite{Coleman:1977py} in
the semiclassical case. We illustrate this fact by a numerical
example.

We end with a set of appendices.


\section{Scalar model in the large N limit - A brief review}\label{useful_formulas}

Let us consider an $O(N)$ symmetric Euclidean action for an $N$ -
component scalar field $\vec{\phi}$ in three space-time dimensions

\begin{equation}
S\left( \vec{\phi}\right) =\int \left[ \frac{1}{2}\left(
\partial _{\mu }\vec{\phi}\right) ^{2}+NU\left( \frac{
\vec{\phi}^{2}}{N}\right) \right] d^{3}x \, .\label{scalaraction}
\end{equation}
The potential has a Taylor expansion of the form \be U\left(
\frac{\vec{\phi}^{2}}{N}\right)
=\sum_{n=1}^{\infty}\frac{g_{2n}}{2n}\left(
\frac{\vec{\phi}^{2}}{N}\right) ^{n} \, ,
\label{ExampleOfPotential}\ee with $g_{2n}$ kept fixed as
$N\to\infty $ (in the large $N$ limit any such potential is
renormalizable \cite{Parisi:1975im}). We describe a way to find the
generating functional of this model. It is given by

\begin{equation}
Z\left[ \vec{J}\right] =\int D\vec{\phi}\exp \left[ -S(\vec{\phi})-
\int\vec{J}(x)\cdot \vec{\phi }(x)d^{3}x\right]  \, .
\end{equation}
Inserting

\begin{equation}
1\sim\int D\rho \delta (\vec{\phi}^{2}-N\rho)\sim \int D\rho
D\lambda e^{-i\int \frac{\lambda}{2}(\vec{\phi} ^{2}-N\rho )d^{3}x}
\label{identity}
\end{equation}
and integrating over $\vec{\phi}$ one obtains
\begin{equation}
Z\left[ \vec{J}\right]= \int D\rho D\lambda\,
 e^{-N S_{eff}(\rho,\,\la) }e^{\frac{1}{2}\int \vec{J} (x)\left(
-\square+i\lambda\right)_{xy}^{-1}\vec{J}(y)d^{3}xd^{3}y}  \, ,
\label{partition}
\end{equation}
where
\begin {equation}
 S_{eff}(\rho,\la)=\frac{1}{2}\int\left[2U(\rho) -i\lambda\rho\right]d^{3}x
 +\frac{1}{2}Tr\ln \left( -\square +i\lambda \right)  \, .
 \label{Seff}
\end{equation}

When $N$ is large, the last form suggests using the saddle point
method to calculate the integral. The two saddle point equations,
obtained by varying the auxiliary fields $\rho$ and $\lambda $
are\footnote{We use here the definition $Tr=\int d^3x\,tr$}

\begin{equation}
2U'(\rho)=i\lambda,\qquad
\rho=tr\frac{1}{-\square+i\lambda}=-\frac{\sqrt{i\lambda}}{4\pi}\equiv-\frac{m}{4\pi}
\, , \label{scalar_gap}
\end{equation}
where $m$ will assume a role of a mass. Moreover, we have used the
dimensional regularization procedure in order to define the
divergent loop. One can also define an effective potential whose
minimum fixes a value of the mass
\begin{equation}
\frac{U_{eff}}{N}=\frac{m^{3}}{24\pi}+\sum\limits_{n=1}^{\infty}\frac{g_{2n}}{2n}\left(-\frac{m}{4\pi}\right)^n
\label{eff_potential2}.
\end{equation}

In the above analysis and in what follows we take into account the
fact that there is no spontaneous breaking of the $O(N)$ symmetry.
The complete vacuum energy \cite{Moshe:2003xn, Bardeen:1983rv} which
accounts for $O(N)$ broken phase is given by
 \begin{equation}
 \frac{U_{eff}}{N}=\frac{m^{3}}{24\pi}+\sum\limits_{n=1}^{\infty}\frac{g_{2n}}{2n}\left({\vec\phi_c^{\,2} \over N}-\frac{m}{4\pi}\right)^n
  \, ,\label{O(N)broken_eff_potential}
 \end{equation}
where $\vec\phi_c$ is the would be expectation value of $\vec\phi$.

When one considers only relevant and marginal terms up to $\phi^6$,
then the potential is stabilized for $0\leq g_6\leq(4\pi)^2$. It is
unstable for $g_6<0$, the instability implied for $g_6>g_c$ is
discussed in the next section. If only $g_6$ is present the theory
is conformal and possesses two $O(N)$ invariant phases, one with
$g_6<(4\pi)^2$ and $m=0$, with $m$ being the mass, and another with
$g_6=(4\pi)^2$ and arbitrary $m$.

In the following sections we generalize the saddle point equations
(\ref{scalar_gap}) to the case when the fields $\rho$ and $\lambda$
are not constant.


\section{Case of non-constant fields - general approach}\label{NonConstantFields}
In this section we introduce a set of equations that govern
evolutionary processes in the theory with Euclidean action given in
(\ref{scalaraction}) in the large $N$ limit. These equations are \be
2U'\bigl(\rho(x)\bigr)=i\la(x),\qquad \rho(x)=G_{reg}(x,x),\qquad
\Bigl(-\Box_x+i\la(x)\Bigr)\,G(x,y)=\delta(x-y) \,
.\label{FullSetOfEquations}\ee Here $G(x,y)$ is the Green's function
and $G_{reg}(x,x)$ is the regularized Green's function \be
G_{reg}(x,x)=\lim\limits_{x\to
y}\Bigl(G(x,y)-\frac{1}{4\pi|x-y|}\Bigr) \,
.\label{RegularizedGreenFunction}\ee

In order to derive these equations recall that, as described in
section \ref{useful_formulas}, any constant solution of the theory
(\ref{scalaraction}) in the large $N$ limit is a solution of the gap
equations (\ref{scalar_gap}) \be 2U'(\rho)=i\la,\qquad
\rho=tr\frac{1}{-\Box+i\la}\ee The trace in the second equation is
$tr\frac{1}{-\Box+i\la}=G(x,x)$, where $G(x,y)$ is a propagator of a
scalar field in a background $i\la$ which solves the last equation
in (\ref{FullSetOfEquations}). The regularization required for the
divergent $G(x,x)$ leads to (\ref{RegularizedGreenFunction}). The
above equations remain valid also for non-constant fields. Any
solution of these equations will describe some process in the
theory. Constant solutions described in section
\ref{useful_formulas} are examples of solutions, but there are many
others. Some examples will be given in what follows.

The main difficulty in finding solutions stems from the fact that
the last equation in (\ref{FullSetOfEquations}) involves computing a
Green's function of the operator with an unknown function $\la(x)$.
Since, to the best of our knowledge, there is no closed expression
for such a Green's function, we shall consider each case separately.

Yet another approach is suggested in \cite{Bardeen:1983st,
Bardeen:1986td}, where the $O(N) ~ g_4 \phi^4$ theory in four
dimensions is explored in a large-$N$ limit. In particular, the
effective action and the corresponding gap equations are derived
within the subspace of slowly varying functions $\lambda(x)$ for
which the inequality $|(\partial \la)^2/ \la^3|<<1$ holds. In
Appendix \ref{SeffDerivation} we build upon this idea to construct
also an effective action in this approximation and analyze different
scenarios of dynamical evolution within it.


\section{Time-dependent solution in the conformal
case}\label{ConformalCase}

In the first case we consider the effective potential
(\ref{ExampleOfPotential}) which contains only $g_6$ - the
dimensionless coefficient, the bare potential is thus
 \be
 U=\frac{g_6}{6\,N^2}\vec{\phi}\,^6 \, , \label{ClassicalPotential}
 \ee
whereas the full quantum effective potential according to
(\ref{eff_potential2}) is \be
U_{eff}(\rho)=\frac{g_6-g_c}{6}\rho^3\label{Ueff_g6},\ee where, as
before, $\rho=\langle\vec{\phi}\,^2\rangle/N$ and $g_c$ is the
critical value of $g_6$ \be g_c=(4\pi)^2\label{DefinitionOfgc}.\ee
As mentioned in section \ref{useful_formulas}, the theory is scale
invariant also quantum mechanically and possesses two different
phases: one with unbroken scale invariance for $g_6<g_c$, and the
other with spontaneously broken scale invariance and a spontaneously
generated mass for $g_6=g_c$. The behavior of the system depends on
the sign of $g_6$ in the classical case and in addition on whether
$g_6$ is bigger or smaller than $g_c$ in the quantum case. Note that
from (\ref{scalar_gap}) it follows that relevant values of $\rho$
after a regularization are negative. The quantum potential for both
signs of $g_6-g_c$ and some types of motion in it that will be
considered below are shown in figure \ref{g6PotentialFigure}. This
potential plays a role in various attempts to obtain a holographic
dual to a gravitational system which may be suspect of exhibiting a
Big Crunch \cite{Hertog:2004rz} .

The unstable potential resulting for $g_6>g_c$  can be misleading.
In that case the UV cutoff $\Lambda$ can not be removed, a UV
completion is needed to have a cutoff independent theory. In this
paper we check what would have been the time scales involved if one
insisted to keep the unstable potential as is, with neither a cutoff
nor a UV completion. This is done to learn about the properties of
such unstable potentials when they actually arise.

\EPSFIGURE{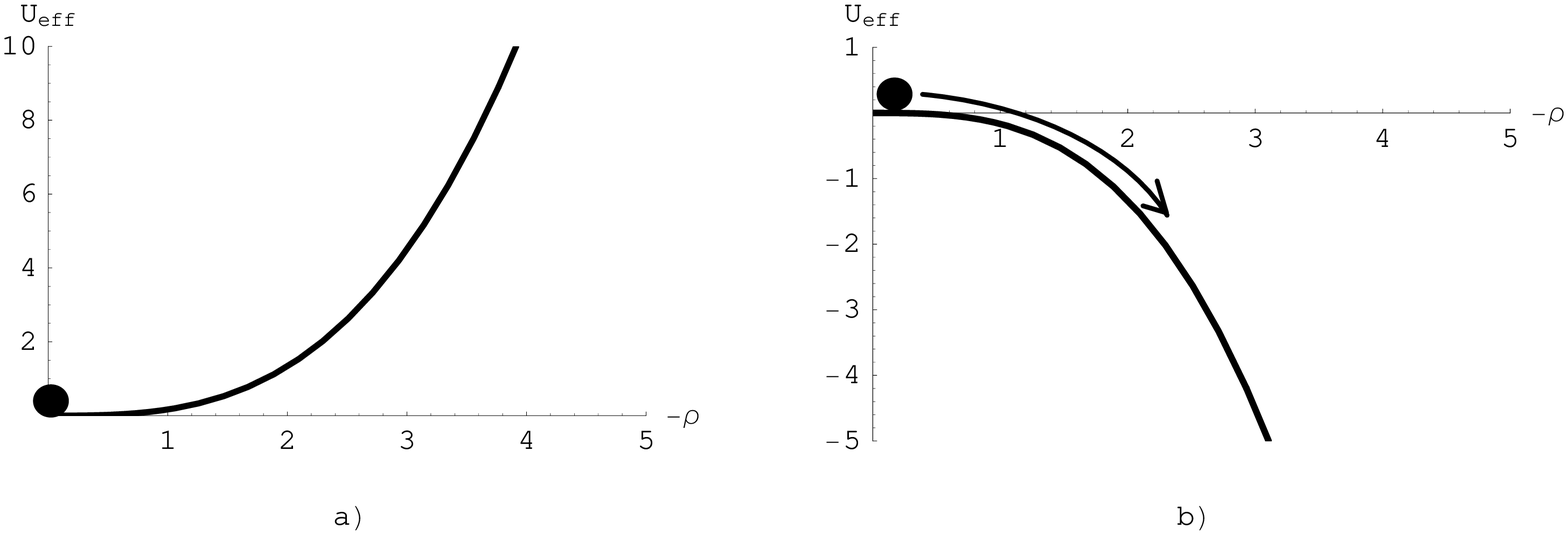,width=15 cm}{a) Effective potential for
$g_6=g_c-1$, b) Effective potential for $g_6=g_c+
1$.\label{g6PotentialFigure}}

In this section we present a solution to equations
(\ref{FullSetOfEquations}) in the scale invariant phase. We consider
a situation when all fields depend only on one coordinate out of
three, and we choose it to be a time $t$. The scale invariance of
the potential dictates the asymptotic form of the time-dependent
solution. Initial conditions such as the initial time $t_0$ and the
initial value of the field at time $t_0$ break scale invariance and
are thus reflected in the solution as well. The solution will
require two integration constants.

The solution that we find is actually the classical motion described
by the full quantum effective action $\Gamma(\vec{\phi})$. We
compare this solution with its classical counterpart, by which we
mean a classical motion in the quantum effective potential
(\ref{Ueff_g6}) with all corrections that include derivatives of the
field neglected. Since $\rho\sim\phi^2$ this last case looks as a
purely classical problem of the motion in the potential $\phi^6$. In
what follows we will call this motion a classical motion, in
contrast with the motion described by the full action
$\Gamma(\vec{\phi})$, which we will call the quantum one.

Here we study a solution for which the classical energy vanishes.
This initial condition does not break scale invariance. However, a
second initial condition, for example, the value of the field
$\phi_0$ at the time $t_0$, does introduce a scale. We compute the
time evolution of the field for both the classical and quantum cases
for zero energy and show that although qualitatively the two cases
are similar there is a quantitative difference: the quantum rolling
is faster. If one considers a time which takes the field to get from
a given initial value to infinity in a way that the total energy
vanishes then the precise relation between these times in the
classical and quantum cases is \be\Delta t_q=\frac{\Delta
t_{cl}}{\sqrt 3}\ee A solution for the non vanishing energy in the
classical case is presented in Appendix
\ref{ArbitraryEnergyAppendix}.

\subsection{The classical case}

We consider here a classical motion of the field in the effective
potential (\ref{Ueff_g6}). We start with the following Lagrangian
 \be
 L= \frac{1}{2}\left(
 \partial _{\mu }\vec{\phi}\right) ^{2}- \frac{a}{6 N^2} \left(
 \vec{\phi}^{2}\right)^3 \, ,
 \ee
where $a$ is an arbitrary constant. For time-dependent and
space-independent solutions, the Lagrangian is
 \be
 L=\frac12\dot{\vec{\phi}}^{\,2}-\frac{a}{6 N^2} \left(
 \vec{\phi}^{2}\right)^3 \, ,
 \ee
where dot means a derivative w.r.t. $t$. The point
${\vec\phi}^{\,2}=0$ in the space of fields corresponds to the
extremum of the potential. In what follows we mostly consider an
inverted conformal potential and investigate those classical
solutions with vanishing total energy, which either asymptotically
tend to ${\vec\phi}^{\,2}=0$ or increase indefinitely. Such
solutions are closely related to the exact quantum ones which we are
going to find in the next subsection, in a sense that for both cases
(classical and quantum) the total energy is zero, the system
approaches the equilibrium point ${\vec\phi}^{\,2}=0$ as $t
\rightarrow \pm\infty$ and the field's magnitude diverges after a
finite time.

The general ansatz for any classical solution of the considered
problem is given by
 \be \vec\phi=\sqrt{N}\phi(x)\hat{n} \, , \ee
where $\phi(x)$ is an arbitrary function and $\hat{n}$ is a unit
vector in the space of fields. For the class of solutions which we
are interested in, $\phi(x)$ can be determined by the scaling
arguments (see below). Moreover, one can show that $\hat{n}=const$
within the subspace of solutions under consideration. Substituting
the above ansatz into the Lagrangian and bearing in mind that
$\hat{n}=const$, one obtains
 \be
 L=N\Bigl(\frac12\dot{\phi}^2-\frac{a}{6}\phi^6\Bigr) \, .
 \label{ClassicalLagrangian}
 \ee
The corresponding EOM is \be \ddot{\phi}=-a\phi^5 \, .\ee By scaling
arguments the solution of vanishing energy should be of the form
 \be
 \phi=\frac{\gamma}{\sqrt{\pm(t-t_{div})}} ~ , \qquad
 \rho_{cl}=\pm\frac{\gamma^2}{t-t_{div}} ~ ,
 \label{ClassicalAnzatzForSolution}
 \ee
where $t_{div}$ is a time when the field may diverge and $\rho_{cl}$
is an analog of $\rho$, which is $\vec{\phi}^{2}/N$. For a positive
value of the coefficient $a$ of the $\phi^6$ potential the only zero
energy solution is that in which the field is stuck statically at
the minimum of the potential. To obtain truly time-dependent
solutions with zero energy one needs to study the case of the
potential unbounded from below obtained for $a<0$. In this case the
EOM determines the value of $\gamma$ to be
\be\gamma=\sqrt[4]{-\frac{3}{4a}} ~ .\label{GammaValue}\ee Also in
this case, there are two different kinds of time-dependent solutions
with zero energy: the field can either escape towards infinity or
asymptotically tend to the maximum of the potential. We will
consider the first case, which implies that in the solution
(\ref{ClassicalAnzatzForSolution}) there should be $t<t_{div}$ and
the sign should be minus. Insisting on placing the field at the
maximum of the potential at time $t_0$ will result with a static
solution remaining there. In order to obtain a time-dependent
solution we can set either $\phi_0$ or $\dot{\phi_0}$ to be
non-zero. The one determines the other through
\be\gamma=\sqrt{\frac{\phi_0^3}{2\dot{\phi}_0}}  \, .\ee With these
initial conditions the solution is
 \be
 \phi=\frac{\gamma}{\sqrt{t_0-t+\frac{\gamma^2}{\phi_0^2}}}
 ~ ,\qquad
 \rho_{cl}=\frac{\gamma^2}{t_0-t+\frac{\gamma^2}{\phi_0^2}} ~ ,
 \ee
with $\gamma$ as in (\ref{GammaValue}). From here one reads off that
\be t_{div}=t_0+\frac{\gamma^2}{\phi_0^2}  ~ . \ee That means that
the field starting at a finite value $\phi_0$ at time $t_0$ reaches
an infinite value after a finite time $\Delta t_{cl}$ which is
 \be
 \Delta t_{cl}=\frac{1}{2\phi_0^2}\sqrt{-\frac{3}{a}} ~ .
 \label{ClassicalGeneralRollingTime}
 \ee
In particular, for the potential (\ref{Ueff_g6}) this time is
 \be
 \Delta t_{cl}=\frac{1}{2\rho_0}\sqrt{\frac{3}{g_6-g_c}} ~ .
 \label{ClassicalRollingTime}
 \ee
This case can be interpreted as a calculating the divergence time in
an exact quantum effective action in which, however, no account was
taken yet of the time derivatives of the fields beyond the standard
kinetic term. We next turn to consider the effect of this time
dependence.

\subsection{The full quantum solution}

Here we consider the full quantum evolution with all time derivative
corrections included. As in the previous section, we consider a
process with vanishing energy. Similarly to the classical case, the
scale invariance fixes the time dependence of the functions
\be\rho(t)=\frac{\bar{\al}_L}{t-t_{div}}~,\qquad
i\la(t)=\frac{\al_L}{(t-t_{div})^2}~.\label{Guess}\ee where the
subscript $L$ stands for ``Lorentzian." In this section we compute
the coefficients $\bar{\al}_L$ and $\al_L$. We show that in addition
to the trivial solution $\bar{\al}_L=\al_L=0$ there exists in the
case $g_6>g_c$ another one
 \be
 \bar{\al}_L=\frac{1}{2\,\sqrt{g_6-g_c}}~,\qquad \al_L=\frac14\,
 \frac{g_6}{g_6-g_c}~.\label{ConformalRollingResults}
 \ee
If $g_6>g_c$ then the quantum potential is unbounded from below in
the sense described before, and the nontrivial solution represents a
rolling from the maximum of the potential towards infinity. However,
if $0<g_6<g_c$ then the potential has a minimum at $\rho=0$ with
vanishing energy. Since we consider only solutions with zero energy,
we find only a trivial one, corresponding to a field at the bottom
of the potential well.

From this one can compute, as for the classical case, a rolling time
$\Delta t_q$, the time it takes the field to roll from a given value
of the field $\rho_0$ to an infinite value, according to the
solution (\ref{Guess}) with coefficients
(\ref{ConformalRollingResults}); this time turns out to be \be\Delta
t_q=\frac{1}{2\,\rho_0\sqrt{g_6-g_c}}~.\ee One can compare this
rolling time with a similar time $\Delta t_{cl}$ for a rolling in
the effective potential (\ref{Ueff_g6}) with $\rho$ replaced by
$\phi^2$, as explained in the previous section. The result is \be
\Delta t_q=\frac{\Delta t_{cl}}{\sqrt 3}~,\ee the quantum rolling is
faster than the corresponding classical one. It should be stressed
that there is no reason for the two times to be equal, since the
effective potential captures only those terms in the full quantum
effective action which do not involve time derivatives of the field
beyond the canonical kinetic term; those terms do however appear in
the case of non-constant solutions. We will see another example of
this phenomenon in the next section, where we consider a rolling in
a general renormalizable potential.

We start all computations in the Euclidean signature. The scale
invariant ansatz for the functions $\rho$ and $i\la$ is \be
\rho(\tau)=\frac{\bar{\al}_E}{\tau-\tau_{div}}~,\qquad
i\la(\tau)=\frac{\al_E}{(\tau-\tau_{div})^2}~,\label{ro}\ee where
$\tau $ is a Euclidean time and $\tau_{div}$ is a Euclidean time
instance at which the field diverges. The subscript $E$ emphasizes
that the work in the Euclidean space-time. In what follows we will
choose $\tau_{div}$ to be zero. This choice fixes a time translation
symmetry of the problem.

The Lorentzian motion is recovered by replacing $\tau$ by $i\,t$.
This, in turn, leads to the expression (\ref{Guess}) for $\rho$ with
$\bar\al_L=-i\,\bar\al_E$. The field $\rho$ is real if $\bar\al_E$
is imaginary. Since, as explained in section
\ref{NonConstantFields}, $\rho$ is a value of the regularized
Green's function with coincident points, the full Green's function
must be complex.

The next step is to check if there is a non real-valued Green's
function of the operator $-\Box+\al_E/\tau^2$. There are two
different cases: 1) $\al_E<-1/4$, 2) $\al_E>-1/4$. The Green's
functions for both cases are derived in Appendix
\ref{GreenFunctiong6}.

 In the case $\al_E>-1/4$ the Green's function is \be
 G(\textbf{r},{\tau},{\tau}_0)=\frac{\sqrt{|{\tau}|\,|{\tau}_0|}}{2\pi}
 \,\frac{1}{W_+\,W_-}\left(\frac{W_+-W_-}{W_++W_-}\right)^{\beta}~,\label{GreenFunctiong6Expression} \ee
 where \be \bt=\frac{\sqrt{1+4\al_E}}{2}~,\label{beta}\ee
and the Euclidean times $\tau$ and $\tau_0$ must be of the same
sign. This Green's function is real and therefore does not lead to a
real-valued field $\rho$. This case corresponds to the potential
$\phi^6$ with a positive coefficient, which is bounded from below.
In this case the only solution with zero energy is the trivial one
$\rho=0$.

Turn now to the case $\al_E<-1/4$, which, as will be shown in course
of the derivation, corresponds to the potential unbounded from
below. In this case $\beta$ in (\ref{beta}) becomes imaginary and
the Green's function (\ref{GreenFunctiong6Expression}) becomes
complex-valued.\footnote{As explained in Appendix
\ref{GreenFunctiong6}, in this case there is also another
real-valued Green's function. It describes a $\rho=0$ solution in
the potential unbounded from below.} The short distance expansion of
this function (for $\textbf{r}\to 0$ and ${\tau}\to {\tau}_0$) is\be
G(\textbf{r},{\tau},{\tau}_0)\simeq\frac{1}{4\pi\sqrt{\textbf{r}^2+(|{\tau}|-|{\tau}_0|)^2}}-\frac{\beta}{4\pi
|{\tau}_0|}~.\ee The first term is a singularity of precisely the
form anticipated in (\ref{RegularizedGreenFunction}), and it is
canceled in the regularized Green's function, which therefore is \be
G_{reg}({\tau},{\tau})=-\frac{\beta}{4\pi
|{\tau}|}\equiv-i\frac{\sqrt{|1+4\al_E|}}{8\pi
|{\tau}|}\label{GreenFunctiong6Regularized}~.\ee Since according to
(\ref{FullSetOfEquations}) $G_{reg}(\tau,\tau)\equiv\rho(\tau)$ we
get \be
\bar{\al}_E=-i\frac{\sqrt{|1+4\al_E|}}{8\pi}\,\textrm{sign}\,\tau
~.\label{AlphaBar}\ee The value is imaginary, and therefore after
the Wick rotation the solution is real. The sign of $\tau$ which
appears in this expression reflects the fact that, as in the
classical case, positive or negative $\tau$'s describe the field
that either runs to infinity or tends asymptotically to zero. From
the first equation in (\ref{FullSetOfEquations}), which in this case
is \be g_6\,\rho^2(\tau)=i\la(\tau) ~,\label{g6FirstEquation}\ee we
conclude that \be \al_E=\frac14\, \frac{g_6}{g_c-g_6}
~.\label{AlphaVSg6}\ee  We see that $\al_E>-1/4$ holds for
$g_6>g_c$, and this is precisely the case when the effective
potential is unbounded from below and one expects to have a rolling
solution. The coefficient $\bar\alpha_E$ defined in (\ref{AlphaBar})
is \be \bar{\al}_E=-\frac{i}{2\sqrt{g_6-g_c}}\,\textrm{sign}\,\tau
~.\label{QuantumSolution}\ee

Wick rotate now to the Lorentzian signature, $\tau=i\,t$. The
Lorentzian solution is the analytic continuation of our result \be
G_{reg}(t,t)\equiv\rho(t)=-\frac{\beta}{4\pi
\,i|t|}\equiv-\frac{\sqrt{|1+4\al_E|}}{8\pi |t|} ~.\ee We see that
$\rho$ is negative, as necessary. The evolution of the field is
either a rolling from the top of the effective potential, which has
no minimum in this case, for $t<0$, or a running towards zero for
$t>0$. We will concentrate on the case of rolling. Plugging in the
value of $\al_E$ from (\ref{AlphaVSg6}) and taking $t<0$ we obtain
\be \rho(t)=\frac{1}{2\,t\,\sqrt{g_6-g_c}},\qquad t<0 ~.\ee This,
after restoring $t_{div}$, leads to the following final form of the
rolling solution
 \be
 \rho(t)=\frac{\bar{\al}_L}{t-t_{div}} ~,\qquad
 i\la(t)=\frac{\al_L}{(t-t_{div})^2} ~,
 \ee
where \be\bar{\al}_L=\frac{1}{2\,\sqrt{g_6-g_c}} ~,\qquad
\al_L=\frac14\, \frac{g_6}{g_6-g_c} ~,\ee and we have restored the
divergence time $t_{div}$.

If $g_6$ is sufficiently close to $g_c$, then all the fields
involved in the problem are slowly varying functions of time and the
approximation suggested in \cite{Bardeen:1983st} is applicable. We
build upon this approximation in Appendix \ref{SeffDerivation} in
order to validate the exact results obtained here. We find a full
agreement.

Moreover, we want to compare these results with the classical ones
of the previous section. In order to do it we consider the quantum
process which is analogous to the process considered there: the
field starts at certain value $\rho_0$ at the initial time $t_0$ and
reaches the infinity after some time $\Delta t_q$. This rolling time
is \be\Delta t_q=\frac{1}{2\rho_0}\frac{1}{\sqrt{g_6-g_c}} ~.\ee
Comparing this rolling time with its classical counterpart
(\ref{ClassicalRollingTime}) we see that the quantum rolling is
faster \be \Delta t_q=\frac{\Delta t_{cl}}{\sqrt 3} ~.\ee As
explained above, the two times need not be equal because in
calculating $\Delta t_q$ one takes into account terms in the
effective action that contain derivatives of the field (in addition
to the usual kinetic term). These terms are not taken into account
in the classical rolling.

In Appendix \ref{QMrolling} we show that one can get a feeling
whether quantum effects tend to accelerate or decelerate a rolling.
The leading order in $\hbar$ correction to the force is $- \sigma^2
U'''(\phi)/2$, where $\sigma$ is the standard deviation. In
particular, in our case according to equation
(\ref{ClassicalLagrangian}) the correction is $ - 10\,a\,\sigma^2
\phi^3$. We see that the classical force $-a\,\phi^5$ and the
quantum correction are of the same sign, and therefore the quantum
correction increases the force. So indeed the quantum rolling should
be faster. In the computation we have determined the precise
relation.

\subsection{Energy conservation}

For completeness we illustrate the energy conservation for the
process under consideration. The energy-momentum tensor in the
Euclidean signature is given here by
 \be
 T_{\mu\nu}(x)=\pd_{\mu}\vec{\phi}\,\pd_{\nu}\vec{\phi}-\delta_{\mu\nu}\Bigl[\frac12(\pd_{\mu}\vec{\phi})^2
 +\frac{g_6}{6N^2}(\vec{\phi}^{\,2})^3\Bigr].\label{TMN}
 \ee
The expectation value of the energy-momentum tensor is calculated by
noting that to leading order in the $1/N$ expansion
  \be <(\vec{\phi}^{\,2})^3>=N^3\rho^3 ~,\label{phi4phi6} \ee
and that
 \be
 <\pd_{\mu}\vec{\phi}(x)\pd_{\nu}\vec{\phi}(y)>=N\frac{\pd^2}{\pd
 x^{\mu}\pd y^{\nu}}G(x,y)~.
 \ee
Using (\ref{GreenFunctiong6Expression}) we get the following
regularized expressions

 \be
 \lim\limits_{\textbf{r},\delta\tau\to 0}
 \frac{\pd^2}{\pd r^2}G(\textbf{r},{\tau},{\tau}_0)
 =\lim\limits_{\textbf{r},\delta\tau\to 0}\frac 1r\frac{\pd}{\pd
 r}G(\textbf{r},{\tau},{\tau}_0)=\frac{\beta(1-\beta^2)}{12\pi\tau^3},\ee
 \be \lim\limits_{\textbf{r},\delta\tau\to 0}\frac{\pd^2}{\pd \tau \pd
 \tau_0}G(\textbf{r},{\tau},{\tau}_0)=\frac{\beta(4\beta^2-7)}{48\pi\tau^3},
 \ee
where $\delta\tau\equiv\tau-\tau_0$.

 In what follows we need to compute only the $T_{00}(x)$ since we are
interested to demonstrate only the energy conservation (recall that
we consider field configurations which depend only on time). First
note that due to rotational and translational invariance in the
$\textbf{r}$-plane one has
 \bea
 <\pd_{i}\vec{\phi}(x)\pd^{i}\vec{\phi}(y)>
 &=& N\frac{\pd^2}{\pd x^{i}\pd y_{i}}G(\textbf{r},{\tau},{\tau}_0)=-N\frac{\pd^2}{\pd x^{i}\pd x_{i}}G(\textbf{r},{\tau},{\tau}_0)
 =-N \triangle G(\textbf{r},{\tau},{\tau}_0) \nonumber \\
  &=&-N  \left( \frac{\pd^2}{\pd r^2} + \frac{\pd}{r \pd r} \right)G(\textbf{r},{\tau},{\tau}_0)\, .
   \label{lap_part_ofEMT}
 \eea
As a result, combining (\ref{ro}),
(\ref{TMN})-(\ref{lap_part_ofEMT}) altogether, we finally obtain
 \be
 \frac{\langle T_{00}\rangle}{N}=\frac{1}{2}\lim_{\textbf{r},\delta\tau\to 0}
  \left( \frac{\pd^2}{\pd \tau \pd \tau_0}\,G(\textbf{r},{\tau},{\tau}_0)
 +\bigl( \frac{\pd^2}{\pd r^2} + \frac{\pd}{r \pd r}
 \bigl)G(\textbf{r},{\tau},{\tau}_0) \right) -\frac{g_6}{6}\rho^3=0 ~.
 \ee
The zero value of the energy of the system is indeed conserved
during the process considered.


\section{Case of small changes of fields}\label{SmallChangesInFields}

\EPSFIGURE{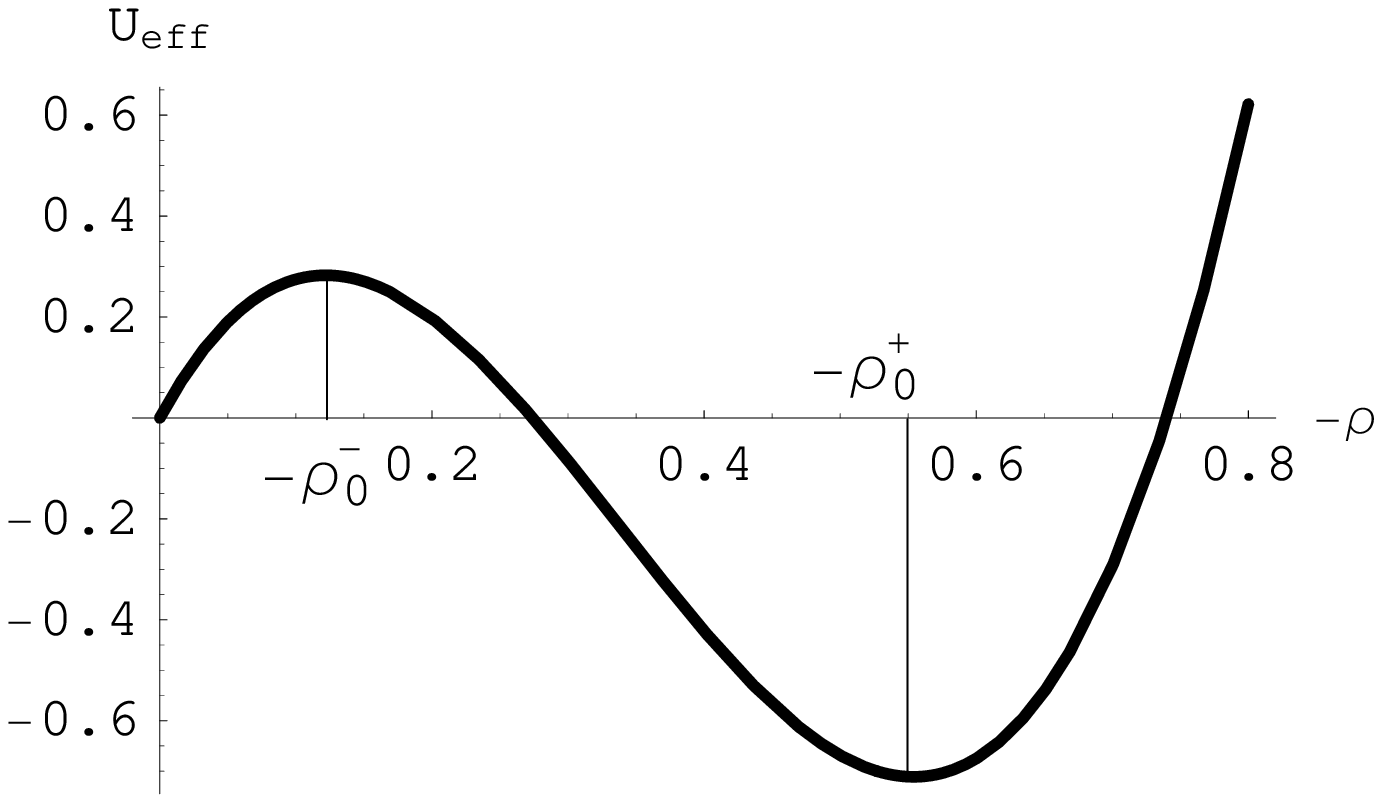,width=6 cm}{Effective potential
(\ref{EffectivePotentialRenormalizable}) for
 $g_2=-10$, $g_4=-100$ and $g_6=10$. The field $\rho$ is negative after
 the renormalization\label{UeffRenormalizableFigure}}

We next consider a case of renormalizable potential with
non-vanishing coefficients $g_2$, $g_4$ and $g_6$ (the restriction
to this particular renormalizable potential is not necessary and can
be omitted without any loss of generality). For our needs we choose
$g_2<0$, $g_4<0$, $0<g_6<g_c$, where $g_c$ is defined in
(\ref{DefinitionOfgc}). The corresponding effective potential
according to (\ref{eff_potential2}) is \be
U_{eff}=\frac{g_2}{2}\rho+\frac{g_4}{4}\rho^2+\frac{g_6-g_c}{6}\rho^3
~. \label{EffectivePotentialRenormalizable}\ee In this expression
the field $\rho$ is constrained to be negative. The potential is
drawn in figure \ref{UeffRenormalizableFigure}. It has a metastable
ground state at the origin and stable minimum away from the origin.

In this section we describe another approach to obtaining
time-dependent solutions. It is valid in an approximation in which
the field does not deviate too much from the extremum of the
potential. The potential (\ref{EffectivePotentialRenormalizable})
possesses two extrema
 \be
 \rho_0^{\pm}=\frac{g_4}{2(g_c-g_6)}\Biggl(1\pm\sqrt{1+\frac{4g_2}{g_4^2}(g_c-g_6)}\Biggr)~.
 \label{rho_pm}
 \ee
These extrema are shown in figure \ref{UeffRenormalizableFigure}.

We study the solution rolling from the top of the potential down to
its minimum and the solution which oscillates around the minimum. We
derive a characteristic time of runaway from the top of the
potential $\rho_0^-$ as well as the frequency of small oscillations
around the true minimum $\rho_0^+$. These will turn out to be
determined by the equations
 \be
 \arctan\frac{\om}{8\pi\rho_0^{\pm}}=\frac{4\pi}{g_4+2g_6\rho_0^{\pm}}\,\om ~.
 \label{OmegaEquation}
 \ee
In the case of oscillations this equation determines the frequency
$\om$, whereas for the rolling it gives the inverse rolling time
$t\sim 1/\om$. Relying on (\ref{rho_pm}) one can show explicitly
that this equation possesses real solutions along with imaginary
ones, depending on the sign before the square root in
(\ref{rho_pm}). We explain their physical meaning below.

We also compare the results with the corresponding classical
counterparts for the same effective potential in the approximation
when the two extrema are close to each other \be
\Bigl|\frac{\rho_0^+-\rho_0^-}{\rho_0^+}\Bigr|\ll1 ~.
\label{approx}\ee We find that within this approximation the
frequency of the quantum oscillations $\omega_q$ is larger than that
of the classical ones $\omega_{cl}$
 \be
 \om_q\simeq \sqrt3\,\om_{cl} ~.
 \ee
The characteristic quantum rolling time for $t_q$ is smaller than
its classical counterpart $t_{cl}$
 \be
 t_q\simeq\frac{t_{cl}}{\sqrt 3} ~.
 \ee

We turn now to the derivation of these results. For the choice of
potential equations (\ref{FullSetOfEquations}) are given by \be
g_2+g_4\rho(x)+g_6\rho^2(x)=i\la(x), \qquad \rho(x)=G_{reg}(x,x),
\qquad \Bigl(-\Box_x+i\la(x)\Bigr)G(x,y)=\delta(x-y)
~.\label{FullSetOfEquationsg2g4g6}\ee Let us assume that the total
change of $\rho$ and $\la$ in the course of dynamical evolution is
small compared to their mean values. This is the case if one would
explore the system during short time intervals. It also holds during
the whole evolution process if one imposes condition (\ref{approx}).

We expand the fields as \bea
 \delta\rho(x)&=& \rho(x)-\rho_0=C\rho_1(x)+C^2\rho_2(x)+..., \nonumber \\
 \delta\la(x)&=& \la(x)-\la_0=C\,\la_1(x)+C^2\,\la_2(x)+...,
 \label{RhoLambdaExpansions}
 \eea
where $\rho_0$ and $i\la_0$ provide a constant solution to
(\ref{FullSetOfEquationsg2g4g6}) and $C$ is an expansion parameter
which is assumed to be small: $|C|\ll 1$. As mentioned in section
\ref{useful_formulas}, in the case of a constant solution at the
minimum of the potential, the value of $i\la$ coincides with a
square of a physical mass, and therefore
 \be i\la_0=m^2 ~. \ee
The last equation in (\ref{FullSetOfEquationsg2g4g6}) can be written
as
 \be
 \Bigl(-\Box_x+m^2+i\dla(x)\Bigr)G(x,y)=\delta(x-y)\, .
 \ee
The solution to this equation can be found by performing a series
expansion
 \begin{multline}
 G(x,y)=G_m(x,y)-\int d^3z\,G_m(x,z)\,i\dla(z)\,G_m(z,y)+\\+\int
 d^3zd^3w\,G_m(x,z)\,i\dla(z)\,G_m(z,w)\,i\dla(w)\,G_m(w,y)+...,
 \label{GreenFunctionSeries}
 \end{multline}
where $G_m(x,y)$ is a propagator of a free massive scalar field
 \be
 G_m(x,y)=\frac{e^{-m|x-y|}}{4\pi|x-y|} ~.
 \ee
One needs to regularize Green's function at coincident points. In
the series expansion (\ref{GreenFunctionSeries}) only the first term
diverges in the limit $x\to y$, and its regularized value is
 \be
 G_{m,reg}(x,x)=\lim\limits_{x\to
 y}\Biggl(\frac{e^{-m|x-y|}}{4\pi|x-y|}-\frac{1}{4\pi|x-y|}\Biggr)=-\frac{m}{4\pi} ~.
 \ee
This result, in particular, shows that $\rho_0=-m/4\pi$ in
accordance with the known result presented in section
\ref{useful_formulas}. Altogether, the expression for $\rho$ becomes
 \begin{multline}
 \rho(x)=-\frac{m}{4\pi}-\int
 d^3z\,G_m^2(x,z)\,i\dla(z)+\\+\int
 d^3zd^3w\,G_m(x,z)\,i\dla(z)\,G_m(z,w)\,i\dla(w)\,G_m(w,x)+...
 \label{RhoSeries}
 \end{multline}
One can solve equations (\ref{FullSetOfEquationsg2g4g6}) order by
order in $C$.
\begin{itemize}
\item{\textit{Order 0.}} To this order equations
(\ref{FullSetOfEquationsg2g4g6}) yield
 \be
 g_2+g_4\rho_0+g_6\rho_0^2=m^2,\qquad
 \rho_0=-\frac{m}{4\pi},\label{Order0}
 \ee
 where, as above, $i\la_0=m^2$. There are two solutions:
 \be
 \rho_0^{\pm}=\frac{g_4}{2(g_c-g_6)}\Biggl(1\pm\sqrt{1+\frac{4g_2}{g_4^2}(g_c-g_6)}\Biggr),\qquad
 i\la_0^{\pm}=g_c\,\rho_0^{\pm} ~.
 \ee

\item{\textit{Order 1.}} In this order one has to compute the
integral in the second term of the RHS in (\ref{RhoSeries}). It
involves a convolution of $G_m^2$ and $i\la_1$, and the result is
 \be
 \rho_1(x)=\frac{1}{4\pi\sB}\arctan\frac{\sB}{8\pi\rho_0}\,i\la_1(x)
 ~.
 \label{Order1}
 \ee

The first equation in (\ref{FullSetOfEquationsg2g4g6}) then gives
 \be
 \Bigl(g_4+2\rho_0 g_6\Bigr)\,\rho_1(x)=\frac{4\pi\sB}{\arctan\frac{\sB}{8\pi\rho_0}}\,\rho_1(x)~.
 \ee
If we look for a solution which depends on a single coordinate, a
Euclidean time $\tau$, this equation simplifies
 \be
 \Bigl(g_4+2\rho_0 g_6\Bigr)\,\rho_1(\tau)=
 \frac{4\pi\,i\pd_{\tau}}{\arctan\frac{i\pd_{\tau}}{8\pi\rho_0}}\,\rho_1(\tau)
 \, ,\label{Rho1Equation}
 \ee
and its form suggests the following solution
 \be
 \rho_1(\tau)=e^{i\om\tau} ~,
 \ee
where $\om$ satisfies equation (\ref{OmegaEquation}).

The result for $\om$ depends on the choice of $\rho_0$: $\om$ is
real for $\rho_0=\rho_0^-$ and imaginary for $\rho_0=\rho_0^+$. This
means the following: $\rho_0^-$ corresponds to a maximum of the
potential and the result describes a runaway solution, whereas
$\rho_0^+$ is a minimum of the potential and in that case $\om$
describes the frequency of the oscillations around it (recall that
the computation was carried out in a Euclidean time and in order to
get a Lorentzian solution one has to Wick rotate the time, which
introduces an additional factor $i$ in the exponents).

The solution for $i\la$ to this order is \be
i\la_1(\tau)=A_1\,e^{i\om\tau}, \qquad
A_1=\frac{4\pi\om}{\arctan\frac{\om}{8\pi\rho_0}}\equiv
g_4+2g_6\,\rho_0 ~.\ee

\end{itemize}

One can continue the perturbative expansion to higher orders. There
will appear corrections both to the frequency of oscillations and to
the shape of the solution. This follows from the fact that if the
amplitude is large enough the field will overshoot the value
$\rho_0^-$ and there will be no oscillations around it.

Let us look more closely at the frequency $\om$ of small
oscillations around $\rho_0^+$. It is determined by equation
(\ref{OmegaEquation}). Assume that the two extrema of the potential
are close to each other, as in equation (\ref{approx}). This
approximation will be valid if the following condition holds
 \be
 \frac{4g_2}{g_4^2}(g_c-g_6)\gtrsim -1 ~.
 \label{CriterionOfValidity}
 \ee
Due to the assumption about the coupling constants ($g_2<0$,
$g_4<0$, $0<g_6<g_c$), this criterion can be satisfied. In this case
$\om$ is small and thus it can be determined by a simpler equation
obtained by expanding the arctangent to the third order
 \be
 \om_q^2\simeq-\frac{3\,g_4^2}{g_c-g_6}\,\sqrt{1+\frac{4g_2}{g_4^2}(g_c-g_6)}
 ~.
 \label{QuantumFreq}
 \ee
This is the Euclidean result, the Lorentzian one will be of opposite
sign. The Euclidean result for $\om_q^2$ is negative for $g_6<g_c$,
and therefore the Lorentzian result will be positive, as expected.
In Appendix \ref{SeffDerivation} we rederive this result in the
framework of another approximation presented in
\cite{Bardeen:1983st}. We find full agreement between the results.

This frequency can be compared to a frequency $\om_{cl}$ of
oscillations of a classical field $\phi$ in an effective potential
(\ref{EffectivePotentialRenormalizable}). Since the potential is
$O(N)$ invariant the most rapid oscillations occur in the radial
direction, thus the maximal value of the classical frequency is
determined by the equation $\om_{cl}^2=U''(\phi_0)$, where
$\phi_0=\sqrt{\rho_0^+}$ is the value of the field $\phi$ at the
extremum. In our approximation the Euclidean result is \be
 \om_{cl}^2\simeq-\frac{\,g_4^2}{g_c-g_6}\,\sqrt{1+\frac{4g_2}{g_4^2}(g_c-g_6)}~.
 \ee Therefore
 \be
 \om_q\simeq \sqrt 3\,\om_{cl}~,
 \ee
and we conclude that the quantum field oscillates faster around the
true minimum than the classical one.

One can carry out a similar computation concerning the
characteristic time of escape from the potential maximum at
$\rho_0^-$. The result is similar: the escape time in the quantum
case is shorter than its classical counterpart by the same factor of
$ \sqrt 3$.

\section{Tunneling}\label{TunnelingSection}
In this section we investigate a possible tunneling in the system.
Following Coleman \cite{Coleman:1977py} we search for solutions
which depend on a Euclidean radius \be
r_E=\sqrt{\vec{r}\,^2+\tau^2},\ee where $\vec{r}$ is a distance in
space and $\tau$ is a Euclidean time. The physical meaning of these
solutions is revealed after the Wick rotation back to the Lorentzian
time. The rotated solution depends on a combination
$r_L=\sqrt{\vec{r}\,^2-t^2}$, where $t$ is a Lorentzian time. Such a
solution describes a spherical wave in the space time. Within the
semiclassical picture of the field evolution after the tunneling (an
expanding bubble with a thin wall) the real and positive values of
$r_L$ correspond to points where the wall has not yet arrived,
whereas imaginary values correspond to those points where it has
already passed.

In what follows we will work in the Euclidean signature and will
suppress the subscript $E$.

Relying on the aforementioned feature \cite{Parisi:1975im} that
three dimensional theory in the large $N$ limit is renormalizable
regardless the shape of the potential, we base our approach on the
reverse engineering method: we will derive the equation for the
instanton, then we will choose a solution which describes a
tunneling in some unknown effective potential, and at the end we
will reconstruct numerically the potential itself. Our consideration
is reliable if the false and the true vacua are close to each other.

Our main results are summarized in figure
\ref{TunnelingResultsFigure1}, where figure
\ref{TunnelingResultsFigure1}a is a plot of a function which
describes a deviation of the solution from its value at the false
vacuum as a function of the Lorentzian radius
 $r=\sqrt{\vec{r}\,^2-t^2}$ for its real values ($r=\infty$ corresponds to the false vacuum itself),
 and figure \ref{TunnelingResultsFigure1}b is a plot of the effective potential with false and true vacua.
\EPSFIGURE{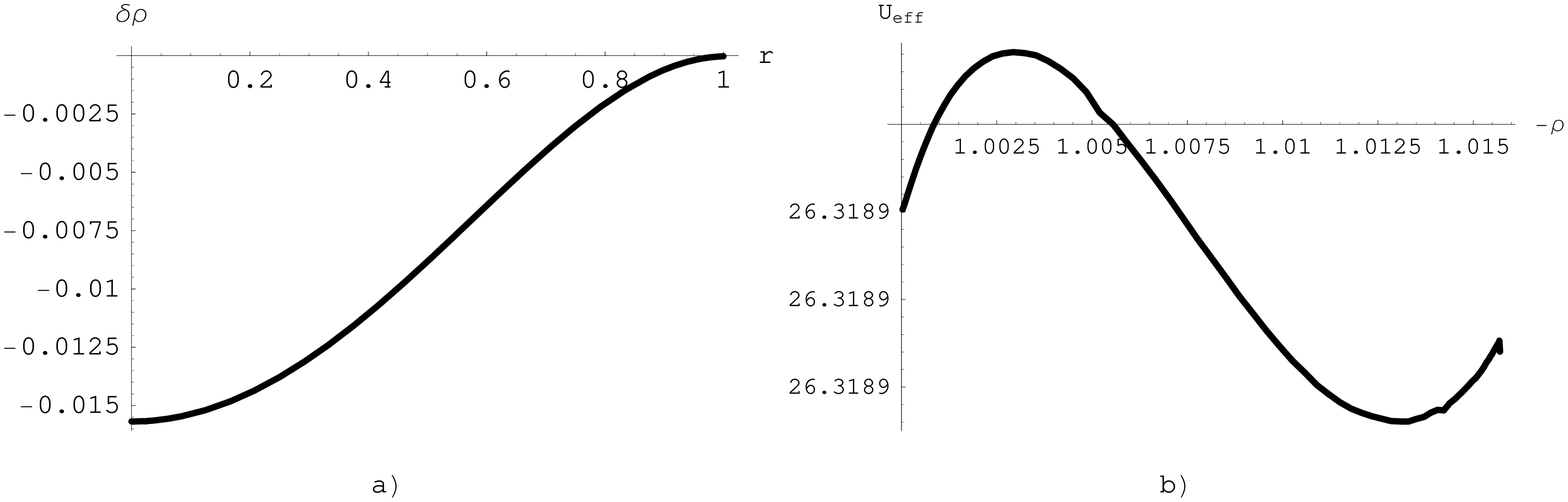,width=15 cm}{a) Function $\drho(r)$
giving a deviation from the false vacuum which is at $r=\infty$, b)
Resulting effective potential $U_{eff}$.
\label{TunnelingResultsFigure1}} We also compute the tunneling
amplitude and compare it to the corresponding semiclassical value.
In both cases the amplitude is $A\simeq\exp(-N\,S)$. In the
semiclassical case $S_{scl}\simeq 4.7$, whereas in the full quantum
computation $S\simeq 3.56$, so the quantum amplitude turns out to be
larger.

We consider a situation when the change in fields during the
tunneling process is small
 \be
 \Bigl|{\rho_t-\rho_f \over \rho_t}\Bigr|<<1 ~,
 \ee
where $\rho_t$ and $\rho_f$ are the values of $\rho$ in the true and
false vacuum respectively. We start with an arbitrary potential
$U(\rho)$ and carry out the solution which presents a combined
numerical description of a tunneling and a subsequent rolling. The
numerical analysis is done relying on the equations which are valid
up to a leading order in the aforementioned small parameter.

Write similarly to (\ref{RhoLambdaExpansions})
 \be
 \rho(x)\simeq-\frac{m}{4\pi}+C\rho_1(x),\qquad
 i\la(x)=m^2+C\,i\la_1(x) ~.
 \ee
To order 0 we will have similarly to (\ref{Order0})
 \be
 2U'(\rho_0)=m^2,\qquad
 \rho_0=-\frac{m}{4\pi}~.
 \ee
The order 1 equation (\ref{Order1}) also remains intact. The two
equations can be combined into a single one
 \be
 2U'\bigl(\rho(x)\bigr)=-(4\pi)^2\rho_0^2+\frac{4\pi\sB}{\arctan\frac{\sB}{8\pi\rho_0}}\,\rho(x) ~.
 \ee
This is a general form of the equation which describes a behavior of
the system to the leading non-trivial order. In the case that we
want to consider, namely, when the field $\rho$ depends only on the
Euclidean radius $r$ the equation simplifies. As we show in Appendix
\ref{SquareRootOfLaplacianInRadialCase}, on functions depending only
on the radius, the operator $\sB$ becomes \be
\sB\,\rho(r)=-\hA\,\rho(r)=-\frac{1}{\pi
r}PV\iii\frac{\bigl(s\,\rho(s)\bigr)'}{s-r}\,ds,\label{DefinitionOfH}\ee
where $PV$ denotes the Cauchy principal value and we assume that the
function $\rho(r)$ has been continued to the region of negative $r$
as an even function\be \rho(-r)=\rho(r) ~.\ee Equation
(\ref{DefinitionOfH}) also serves as a definition of an operator
$\hA$ which will be extensively used in what follows.

From what was said above we arrive at the final form of the equation
for $\rho(r)$
 \be
 2U'\bigl(\rho(r)\bigr)=-(4\pi)^2\rho_0^2+\frac{4\pi\,\hA}
 {\arctan\frac{\hA}{8\pi\rho_0}}\,\rho(r) ~.
 \label{TunnelingEquationRho}
 \ee
This equation should be supplemented by boundary conditions for
$\rho(r)$. Following \cite{Coleman:1977py} we require that $\rho(r)$
go its value at the false vacuum when $r\to\infty$. We denote this
value of $\rho$ by $\rho_0$ and define
 \be \drho(r)=\rho(r)-\rho_0 ~.\ee
The function $\drho(r)$ goes to 0 as $r$ goes to infinity. With this
definition equation (\ref{TunnelingEquationRho}) becomes
 \be
 2U'\bigl(\rho_0+\drho(r)\bigr)-2U'(\rho_0)=
 \frac{4\pi\hA}{\arctan\frac{\hA}{8\pi\rho_0}}\,\drho(r),\qquad
 2U'(\rho_0)=(4\pi)^2\rho_0^2 ~.
 \label{NonLinearTunnelingEquation}
 \ee
Linearized around $\rho_0$
 \be
 U''(\rho_0)\,\drho(r)=
 \frac{2\pi\hA}{\arctan\frac{\hA}{8\pi\rho_0}}\,\drho(r) ~.
 \label{LinearizedTunnelingEquation}
 \ee
Note, that since at $\rho_0$ the potential has a minimum
$U''(\rho_0)$ is positive. However, as shown in Appendix
\ref{SquareRootOfLaplacianInRadialCase}, all eigenvalues of $\hA$
are negative, and, since $\rho_0<0$, it follows that the operator on
RHS of (\ref{LinearizedTunnelingEquation}) is negatively definite.
Therefore there are no solutions to equation
(\ref{LinearizedTunnelingEquation}). Nevertheless, this feature may
not pose a problem since one cannot extrapolate it to the solution
of equation (\ref{NonLinearTunnelingEquation}).

One can rewrite equation (\ref{NonLinearTunnelingEquation}) in the
following form
 \be
 U''(\rho_0)\,\drho(r)+\sigma\bigl(\drho(r)\bigr)=\hB\,\drho(r),\qquad
 \hB=\frac{4\pi\hA}{\arctan\frac{\hA}{8\pi\rho_0}}
 \label{TunnelingEquationWithSigma},
 \ee
where the function $\sigma\bigl(\drho(r)\bigr)$ is proportional to
$\drho^2(r)$. Since for large real $r$ the function $\drho(r)$ goes
to $0$ we conclude that it should become an eigenfunction of $\hB$
at least for very large $r$.

We say that a function $f(r)$ is an  ``asymptotic eigenfunction" of
the operator $\hB$ if there is a number $k$ (an ``asymptotic
eigenvalue") such that $f$ obeys the following requirement
 \be
 \lim\limits_{r\to\infty}f(r)=0,\qquad
 \lim\limits_{r\to\infty}\frac{\hB\,f(r)-k\,f(r)}{f(r)}=0 ~.
 \label{AsymptoticEigenfunctionDefinition}
 \ee

In Appendix \ref{SquareRootOfLaplacianInRadialCase} we show that the
operator $\hB$ possesses many asymptotic eigenfunctions with
positive asymptotic eigenvalues (see (\ref{bB}) and a discussion
there and remember that $\rho_0<0$).  Consider the following
illustrative example.

The construction of the candidate function $\drho$ starts with a
function $f_1(r)$
  \be f_1(r)=\Bigl\{\begin{array}{cc}
                    5r(r^2-1)^2,\qquad &r<1\\
                    0,\qquad &r>1
                 \end{array}\label{f1}
  \ee
This function is plotted in figure
\ref{TunnelingAuxiliaryFunctionsFigure}a. Without any loss of
generality, let us take both the eigenvalue and $\rho_0$ to be equal
$-1$. Then one can compute (see Appendix
\ref{SquareRootOfLaplacianInRadialCase} for details) the functions
$\drho$ and $\hB\drho$
 \bea
 \drho(r)&=&-{2 \over \pi}\int\limits_0^{\infty}\Psi(k)\frac{\arctan
 \frac{k}{8\pi}}{4\pi k+\arctan \frac{k}{8\pi}}\frac{\sin k r}{
 r}\,dk  ~,
 \nonumber \\
 \hB\drho(r)&=& {2 \over \pi} \int\limits_0^{\infty}\Psi(k)\frac{4\pi
 k}{4\pi k+\arctan \frac{k}{8\pi}}\frac{\sin k r}{ r}\,dk ~,
 \label{drhoTunneling}
 \eea
where $\Psi(k)$ is the sine-Fourier transform of $f_1(r)$
 \be
 \Psi(k)=\int_{0}^{\infty}f_1\,(r)\sin(kr)dr=\frac{40k(k^2-15)\cos
 k-120(2k^2-5)\sin k}{k^6} ~.
 \ee
These functions are plotted in figure \ref{TunnelingResultsFigure1}a
and \ref{TunnelingAuxiliaryFunctionsFigure}b. One can also
reconstruct the classical potential from equation
(\ref{NonLinearTunnelingEquation}), and then, using the definition
of the effective potential (\ref{eff_potential2}) \be
U_{eff}(\rho)=U(\rho)-\frac{(4\pi)^2}{6}\rho^3 ~,\ee one can
reconstruct it as well. The result is presented in figure
\ref{TunnelingResultsFigure1}b.
\EPSFIGURE{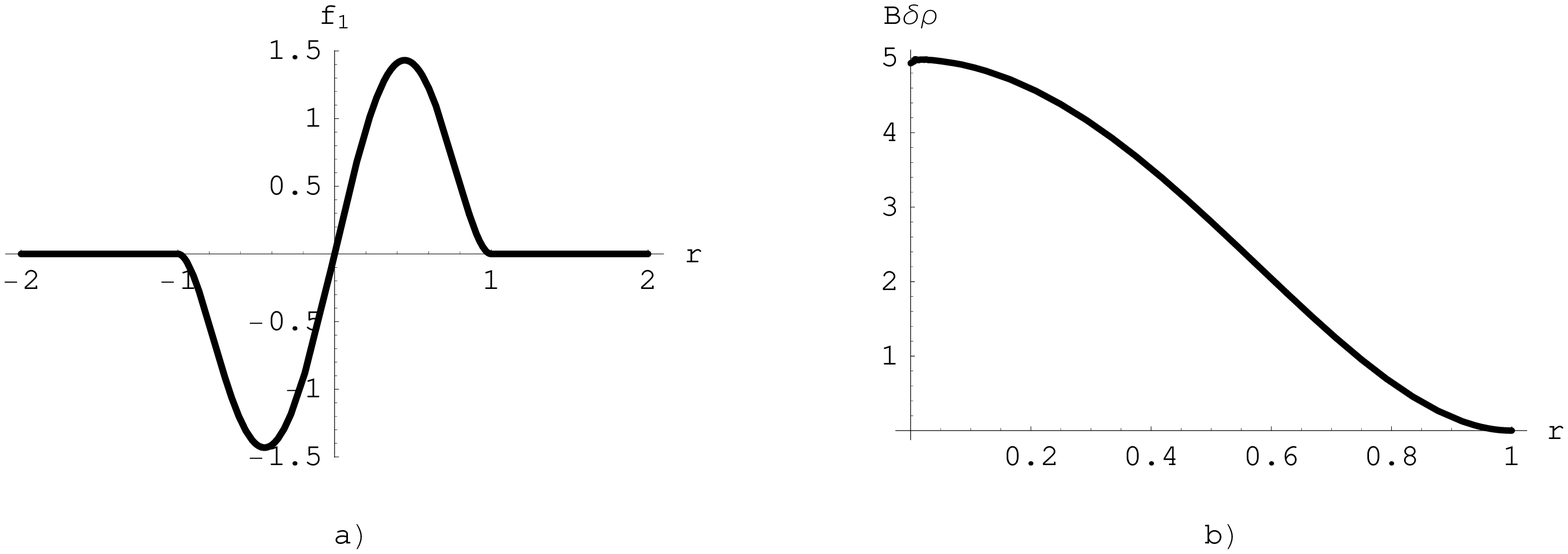,width=15 cm}{a) Function
$f_1(r)$, as defined in (\ref{f1}), b) Function $\hat{B}\drho(r)$.
\label{TunnelingAuxiliaryFunctionsFigure}} We see that the potential
indeed increases from its value at $\rho=1$ (the fact that its
derivative does not vanish at this point is a numerical error, the
numerics is not reliable at this region; the derivative is very
small though). We see also that the effective potential has a
minimum at $\rho\simeq 1.012$ and then grows up. Since
$\rho-\rho_0\ll\rho$ the approximation is valid along the process .

These results have the following interpretation. At $t=0$ the field
tunnels from the false vacuum and a bubble is created. Its shape is
given by the function $\rho(r)$ that we computed, and there are
points in space where the field acquires all values that are covered
by the plot of the potential \ref{TunnelingResultsFigure1}b. Then,
as the time grows, the solution evolves according to the law
$\rho(\sqrt{\vec{r}^2-t^2})$. We see that the initial shape of the
bubble is complicated, contrary to the case of the thin wall
approximation.

The fact that our numerical computation reconstructs the minimum of
the potential and even goes further means the following. In the
semiclassical approximation the computation similar to the one
presented here would reconstruct the potential up to a point which
is just a little beyond the point where the value of the potential
is equal to its value at the false vacuum. That the field at all
gets beyond the turning point (a point where the potential is
precisely equal to its value at the false vacuum) is because of the
friction term in the EOM. If there is no friction term, like in the
case of quantum mechanics, the computation will only be able to
reconstruct the potential up to its turning point. Our computation
gets so far beyond the turning point and even beyond the true vacuum
because the action we work with is very complicated. In particular,
it possesses a non-standard kinetic term, which, in turn, leads to a
non-trivial friction.\footnote{We are grateful to J. Barbon for
discussions on these issues.}

Having computed the potential, one can also evaluate the effective
action on the solution in order to get the tunneling amplitude. In
the case under consideration the action is $S_{eff}\simeq 3.56$, and
the tunneling amplitude is $\exp(-N\,S_{eff})$.

This result can be compared with a prediction of the semiclassical
approximation for the same potential. In that computation one has to
solve a radial Euclidean EOM for a bounce and then to evaluate its
action. The numerical result in this case is $S_{sc}\simeq 4.7$, and
the tunneling amplitude, similarly to the case above, is
$\exp(-N\,S_{eff})$. We see that the full quantum tunneling
amplitude is larger in this case. Once again this falls within the
pattern that taking account of time-dependent effects shortens the
characteristic time of the process.


\section{Conclusions}\label{Conclusion}

We have studied a variety of time-dependent processes. These
included rolling of fields, tunneling  among vacua and their decay.
This was done for a large class of vector $O(N)$ models in three
space time dimensions, among them  the system which is conformal
also quantum mechanically. Using the methods of the large $N$
expansion we were able to take into account the effects of the time
variation of the fields on the exact effective potentials. We have
calculated exact and approximate characteristic time scales of such
processes. In most cases the results were  quantitatively different
than those obtained without considering the time dependency in the
effective potential. A qualitative difference is found when
analyzing the bubble driving the decay of a false vacuum. The bubble
shows thick rather than thin characteristics. For the cases studied
a pattern of accelerated time scales emerged. The next natural step
is to study supersymmetric extensions of such systems and their
coupling to gravity.


\section*{Acknowledgements}
We thank O.Aharony, T.Banks, J.Barbon, S.Elitzur, B.Kol and
N.Seiberg for discussions at various stages of this work. This
research is partially supported by DIP grant H.52, The Einstein
Center at the Hebrew University, the American-Israel Bi-National
Science Foundation, the Israel Science Foundation Center of
Excellence and by the European Union Marie Curie RTN network under
contract MRTN-CT-2004-512194. Two of us (V.A. and M.S.) are also
partially supported by The Israel Science Foundation grant. E.R.
thanks the New High Energy Theory Center of Rutgers University for
the kind hospitality during which parts of this work have been done.


\section*{\Large{Appendices}}

\appendix

\section{Effective action in the region of slowly varying fields}\label{SeffDerivation}
In this appendix we follow \cite{Bardeen:1983st} in order to compute
the effective action which governs the dynamics of the large-$N$
theory within the subspace of slowly varying functions $\lambda(x)$,
for which the inequality $|(\partial \la)^2/ \la^3|<<1$ holds. We
compare the results obtained for the exact conformal case and those
obtained for small field variations. The comparison is done in those
regions where the validity of the approximations overlap.

For simplicity of notation we suppress imaginary unit $i$ in front
of $\lambda$ in what follows, that is $i\lambda\rightarrow \lambda$.

Let us define $\Gamma(\lambda)$ from the Euclidean functional
integral

\begin{eqnarray}
e^{-\Gamma(\lambda)}&=&\int D\vec{\phi}
 \exp \left( -\int d^{\,3}x \left[ \frac{1}{2}\vec{\phi}
 (-\Box + \lambda)\vec{\phi} \right] \right) \nonumber \\
 &=&\exp \left[-\frac{N}{2}Tr \ln (-\Box + \lambda) \right] ~.
\end{eqnarray}
In the regime of aforementioned approximation $\Gamma(\lambda)$ can
be written as a local expansion

\begin{equation}
\Gamma(\lambda)=\int d^{\,3}x \left[ F_{0}(\lambda)
 + F_{1}(\lambda) (\partial_{\mu}\lambda)^2 + \,.\,.\,. \right]
 \label{Gamma} ~,
\end{equation}
where $F_{0}(\lambda)$ and $F_{1}(\lambda)$ are local functions of
$\lambda(x)$. In particular, $F_{0}(\lambda)$ is found by
calculating the $Tr \ln(-\Box + \lambda)$ with a constant $\lambda$.
Up to an infinite constant, which is zero in the dimensional
regularization, we have

\begin{eqnarray}
\frac{N}{2} Tr \ln(-\Box + \lambda) &=& \frac{N}{2}
 \int \frac{d^{\,3}x \, d^{\,3}p}{(2\pi)^3} \ln \left( 1 + \frac{\lambda}{p^{\,2}} \right) \nonumber \\
 &=& -\frac{N}{12\pi} \int d^{\,3}x \, \lambda^{3/2} ~,
\end{eqnarray}
and thus
\begin{equation}
 F_{0}(\lambda) = -\frac{N}{12\pi} \, \lambda^{3/2} ~.
 \label{F0}
\end{equation}
On the other hand, in order to calculate $F_{1}(\lambda)$ one notes
that

\begin{eqnarray}
-\frac{\delta^2\Gamma(\lambda)}{\delta\lambda(x)\delta\lambda(y)}&=&
 \frac{1}{4}\left[ \langle \vec{\phi}^2(x)\vec{\phi}^2(y) \rangle
 - \langle\vec{\phi}^2(x)\rangle \langle\vec{\phi}^2(y)\rangle \right]
 \nonumber \\
 &=& \frac{N}{2} \int \frac{d^{\,3}k}{(2\pi)^3}e^{i k(x-y)} \int \frac{d^{\,3}p}{(2\pi)^3}
 \frac{1}{((p-k)^2+\lambda)(p^2+\lambda)}
 \nonumber \\
 &=& \frac{N}{16 \pi} \int \frac{d^{\,3}k}{(2\pi)^3} \, e^{i k(x-y)}
 \int_{0}^{1} d\alpha \left( \alpha(1-\alpha)k^2 + \lambda
 \right)^{-1/2} ~.
\end{eqnarray}
This can be expanded in the form

\begin{eqnarray}
-\frac{\delta^2\Gamma(\lambda)}{\delta\lambda(x)\delta\lambda(y)}
 &=& \frac{N}{16 \pi} \int \frac{d^{\,3}k}{(2\pi)^3} \, e^{i k(x-y)}
 \left( \lambda^{-1/2}- \frac{k^2}{12\pi}  \lambda^{-3/2} + \textit{O}(k^4)\right)
 \nonumber \\
 &=& \frac{N}{16 \pi}\lambda^{-1/2}\delta(x-y)
 +\frac{N}{192 \pi}\lambda^{-3/2}\Box\delta(x-y)
 \nonumber \\
 &+&\textit{O}(\Box^2\delta(x-y)) ~,
 \label{localGamma}
\end{eqnarray}
where in the second equality we assume that $\lambda$ is constant.
Such an expansion is justified due to the assumed approximation
regime of slowly varying fields. Using equation (\ref{Gamma}) we can
now identify $F_{0}(\lambda)$ and $F_{1}(\lambda)$ with the
constant-$\lambda$ expansion of equation (\ref{localGamma})

\begin{eqnarray}
\frac{\partial^2F_{0}(\lambda)}{\partial^2\lambda}&=&
 -\frac{N}{16\pi}\lambda^{-1/2} ~,
 \nonumber \\
 F_{1}(\lambda)&=& \frac{N}{384 \pi}\lambda^{-3/2} ~.
\end{eqnarray}
Integrating the first of these equations reproduces the result of
(\ref{F0}). Thus, effective action (\ref{Seff}) can be rewritten as
follows

\begin{equation}
S_{eff}(\rho,\lambda)= \int d^{\,3}x \left[
 U(\rho)-\frac{\rho\lambda}{2} - \frac{\lambda^{3/2}}{12\pi} + \frac{\lambda^{-3/2}}{384 \pi}
 (\partial\lambda)^2\right] ~.
\end{equation}
If we now define

\begin{equation}
\psi(x)=\frac{\lambda(x)^{1/4}}{\sqrt{12\pi}} ~,
\end{equation}
then effective action can be written as follows

\begin{equation}
S_{eff}(\rho,\psi)= \int d^{\,3}x \left[
 U(\rho)-\frac{(12\pi)^2}{2}\rho\psi^4 - (12\pi)^2\psi^6 + \frac{1}{2}
 (\partial\psi)^2\right]~.
 \label{eff_action}
\end{equation}

Let us exploit this action in order to illustrate the process of
tunneling and rolling in the case when the potential $U(\rho)$ is
given by
 \be
 U(\rho)={g_2 \over 2} \rho + {g_4 \over 4} \rho^2 ~.
 \ee
Under this assumption the effective action $S_{eff}(\rho,\psi)$
turns out to be quadratic in auxiliary field $\rho$ and thus
integrating it out yields
\begin{equation}
 S_{eff}(\psi)= \int d^{\,3}x \left[
 \frac{1}{2}(\partial\psi)^2- {(12\pi)^4\over 4 g_4}\psi^8- (12\pi)^2\psi^6 + {g_2 \over 2 g_4}(12\pi)^2 \psi^4 \right]~.
\end{equation}
Let us explore the temporal rolling first, that is we consider the
situation when all the fields are time-dependent only. As a result,
the corresponding Lorentzian equation of motion represents a
particle of unit mass moving in a potential
\begin{equation}
 V(\psi)= - {(12\pi)^4\over 4 g_4}\psi^8- (12\pi)^2\psi^6 + {g_2 \over 2 g_4}(12\pi)^2
 \psi^4 ~.
 \label{Veff}
\end{equation}
Thus,
 \begin{equation}
  E=\frac{1}{2}\left({d \psi \over dt}\right)^2 + V(\psi)
 \end{equation}
is a constant of the motion. This can be used to determine the
qualitative features of the solutions by inspection.

As a simple example of rolling, consider the potential shown in
figure \ref{Veff_fig} which corresponds to
 \be
 -\left({g_4 \over 8\pi}\right)^2 < g_2 < 0, \, g_4<0 \, .
 \ee
We are interested to investigate the rolling of the system from the
local maximum of the potential situated at
 \be
 \psi_{-}^2 =-{g_4 \over 96 \pi^2} \left( 1-\sqrt{1+\left({8\pi \over g_4}\right)^2 g_2}
 \right)\, ,
 \ee
down to the false vacuum located at
 \be
 \psi_{+}^2 =-{g_4 \over 96 \pi^2} \left( 1+\sqrt{1+\left({8\pi \over g_4}\right)^2 g_2}
 \right)\, .
 \ee
\EPSFIGURE{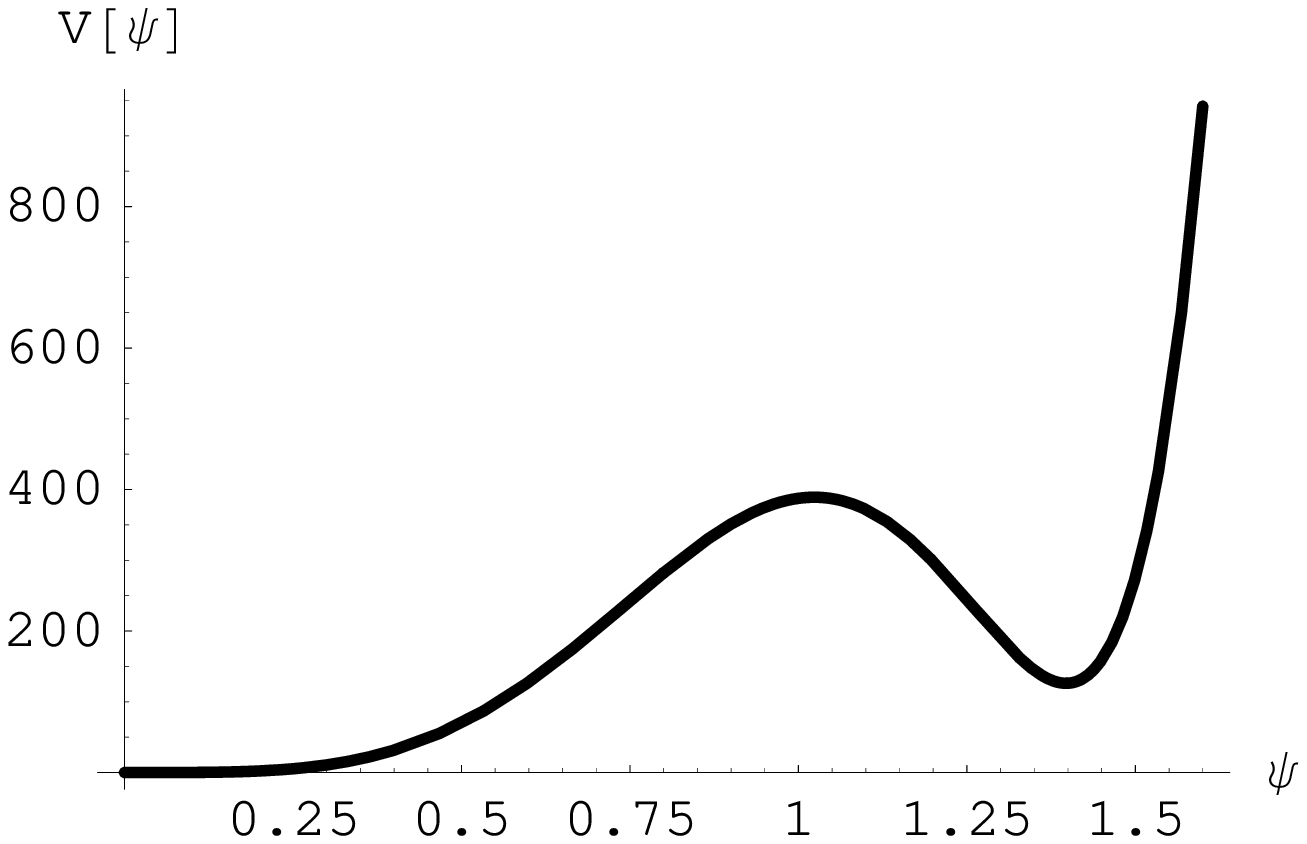,width=6 cm}{The effective potential
(\ref{Veff}) for $g_4=-(12\pi)^2, \, g_2=0.91(g_4/8\pi)^2$.
\label{Veff_fig}} The corresponding runaway and oscillation
frequencies are given respectively by
 \be
 \omega^2_{\pm}={d^2 V \over d\psi^2_{\pm}} ~.
 \ee
In this case, we are dealing with a solution of the equation of
motion with $E=V(\psi_{-})$, whence $\psi$ as a function of $t$ is
given implicitly by
 \be
 t=t_{+} + \int_{\psi_{+}}^{\psi}  {d\psi \over
 \sqrt{2(V(\psi_{-})-V(\psi))}} ~,
 \ee
where $t_{+}$ is an integration constant, the time at which $\psi$
equals $\psi_{+}$.

On the other hand, according to Coleman \cite{Coleman:1977py}, if we
are interested to compute a decay probability per unit time per unit
volume $\Gamma/V$, of the unstable state $\psi_{+}$ due to the
barrier penetration, one must find the bounce $\bar\psi$, a solution
of the Euclidean equations of motion
 \be
 {d^2\bar\psi \over dr^2}+{2 \over r}{d\bar\psi \over dr}=V'(\psi)
 \ee
subject to the following boundary conditions
 \be
 \lim_{r\rightarrow \infty}\bar\psi(r)=\psi_{+} \,~ ,
 ~\bar\psi'(r=0)=0 \, ,
 \ee
where prime denotes derivative with respect to Euclidean radius $r$.
To leading order in $1/N$,
 \be
 \Gamma / V \,\thicksim\, e^{-N\,S_{eff}(\bar\psi)}(1+ \textit{O}(1/N)) \, .
 \ee

In other words, in the limit of large $N$ barrier penetration is
exponentially small and thus, as emphasized in the text, the
dynamics of the system is governed by the rolling processes only.

It is instructive to perform the calculation in the region where the
approximation of slowly varying fields presented above overlaps with
the small field approximation used in the section
\ref{SmallChangesInFields}. Therefore let us assume that condition
(\ref{CriterionOfValidity}) holds. As a result, the two extrema of
the potential plotted on figure \ref{UeffRenormalizableFigure} can
be made arbitrary close to each other, and the field derivatives in
turn, during the evolution of the system from the top of the
potential down to its minimum, become arbitrary small.

As an illustrative example, let us derive the frequency of small
oscillations around the true minimum $\rho_0^+$. Relying on
(\ref{eff_action}) the corresponding EOM are given by
 \bea
 \Box\psi&=&-2(12\pi)^2\rho\psi^3-6(12\pi)^2\psi^5 \, ,
 \nonumber \\
 (12\pi)^2\psi^4&=&g_2+g_4\rho+g_6\rho^2 \, .
 \label{EOM}
  \eea
Since the oscillations are small, we linearize around the true
vacuum
 \bea
 \Box\delta\psi&=&-2(12\pi)^2(3\rho_0^+\psi_0^{+\,2}\delta\psi+\psi_0^{+\,3}\delta\rho)-30(12\pi)^2\psi_0^{+\,4}\delta\psi \, ,
 \nonumber \\
 \delta\rho &=& {4(12\pi)^2\psi_0^{+\,3}\delta\psi\over  g_4+2 g_6\rho_0^+} \, ,
  \eea
where
 \bea
 \psi_0^{+\,2}&=&-{\rho_0^+ \over 3} \, ,
 \nonumber \\
 \delta\rho&=&\rho - \rho_0^+  \, ,
  \nonumber \\
 \delta\psi&=&\psi - \psi_0^+ \, .
 \eea
As a result, the quantum frequency of oscillations is given by
 \be
 \omega_q^2=12g_c\rho_0^{+\,2}\left[ 1- {2 g_c \rho_0^+ \over g_4 + 2 g_6 \rho_0^+}\right]
 \simeq-\frac{3\,g_4^2}{g_c-g_6}\,\sqrt{1+\frac{4g_2}{g_4^2}(g_c-g_6)}
 \, .
 \ee
This result agrees with (\ref{QuantumFreq}) obtained via small field
approximation.

The other case we will consider is that in which only $g_6$, the
dimensionless coefficient, is present. Equations (\ref{EOM}) in this
case combine together and yield
 \be
 {\Box\psi \over \psi^5}=-6(12\pi)^2\left[ 1-\sqrt{{g_c \over g_6}} \right] \, .
 \ee
Since in the region of slowly varying fields $1>> |(\partial \la)^2/
\la^3| \thicksim |\Box\psi/\psi^5|$ one concludes that
 \be
 \delta g = g_6 - g_c << 1 \,
 \ee
in order to justify the approximation. This yields
 \be
 {\Box\psi \over \psi^5}=-27 \, \delta g \, .
  \ee
Therefore according to (\ref{ClassicalAnzatzForSolution}) the
time-dependent and space-independent solution with vanishing energy
is given by
 \be
 \psi=\frac{\sqrt[4]{-3/(4a)}}{\sqrt{\pm(t-t_{div})}}
 \ee
with\footnote{Note that (\ref{ClassicalAnzatzForSolution})
corresponds to the solution of the Lorentzian EOM, whereas
(\ref{EOM}) corresponds to the Euclidean time. This reveals the
origin of an extra minus sign in the expression for $a$.} $a=-27\,
\delta g$.

Altogether
 \be
 \lambda = (12\pi)^2\psi^4 \backsimeq {g_6 \over 4 (g_6 - g_c)}
 {1 \over(t-t_{div})^2} \, ,
 \ee
which coincides with the exact results (\ref{Guess}),
(\ref{ConformalRollingResults}).

\section{Classical motion in the $\phi^6$ potential}\label{ArbitraryEnergyAppendix}
In this appendix we present a solution which describes the classical
evolution of a particle with energy $E$ in the $\phi^6$ potential.
We derive the divergence time as a function of the energy of the
particle.

Consider the following classical Lagrangian \be L=\frac12
\dot{x}^2+\frac a6 x^6 ~,\ee which describes a run-away potential
(for positive $a$). A general solution is \be x(t)=(9\sqrt
3-15)^{1/6}\left(\frac
Ea\right)^{1/6}\sqrt{\frac{1-\textrm{cn}(a^{1/6}E^{1/3}\,\xi,k)}
{2-\sqrt3+\textrm{cn}(a^{1/6}E^{1/3}\,\xi,k)}} ~,\ee where \be
\xi=2^{4/3}3^{1/12}(t-t_0),\qquad k=\frac14(2+\sqrt3) ~,\ee and
$t_0$ and $E$ are integration constants.
\EPSFIGURE{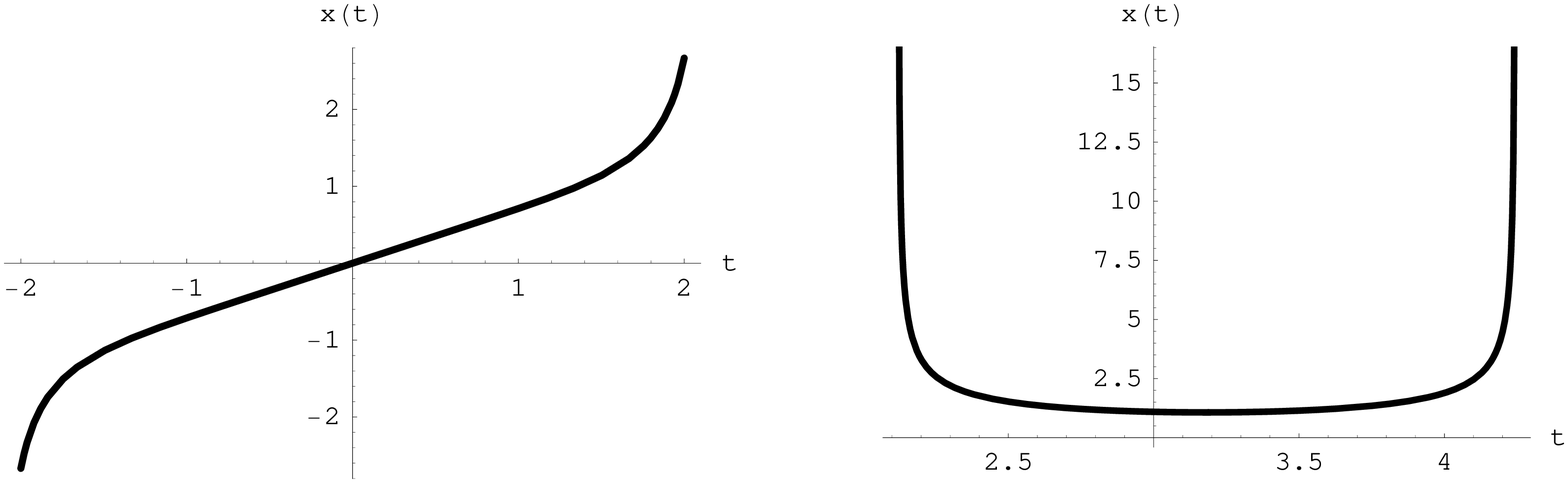,width=12 cm}{Solutions $x(t)$ for
$a=1$ and energies $E=\pm 0.25$. \label{Solutions_Figure}} $E$ is
the total energy and $t_0$ is an arbitrary time. $\textrm{cn}(x,k)$
is the elliptic cosine function. In figure \ref{Solutions_Figure}
there are the solutions for $a=1$ and $E=\pm 0.25$.  We see that the
solution with positive energy passes through 0, whereas the one with
negative energy never reaches 0 (a particle comes from infinity and
bounces back).

If a particle starts at some point $x_0$ at time $t_0$ with energy
$E$ then it gets to infinity after a time
\begin{multline} \Delta
t=\frac{1}{2^{4/3}3^{1/12}a^{1/6}E^{1/3}}
\Biggl(\textrm{cn}^{-1}(\sqrt 3-2,k)-\\-\textrm{cn}^{-1}\Bigr(
\frac{3(3\sqrt3-5)^{1/3}\,E^{1/3}+2^{1/3}3^{1/6}(3-2\sqrt3)\,a^{1/3}\,x_0^2}
{3(3\sqrt3-5)^{1/3}\,E^{1/3}+2^{1/3}3^{2/3}\,a^{1/3}\,x_0^2},k
\Bigr)\Biggr) ~,
\end{multline}
where $\textrm{cn}^{-1}(x,k)$ is the inverse elliptic cosine
function.  This divergence time is plotted as a function of energy
in figure \ref{Divergence_Time_Figure}. We see that the more is the
energy the less is the time that takes the particle to get to
infinity. For small energies this divergence time is \be \Delta
t=\frac{1}{2x_0^2}\sqrt{\frac{3}{a}}-0.67\frac{E}{x_0^8a^{3/2}}
~,\ee and for vanishing energy one recovers
(\ref{ClassicalGeneralRollingTime}).\EPSFIGURE{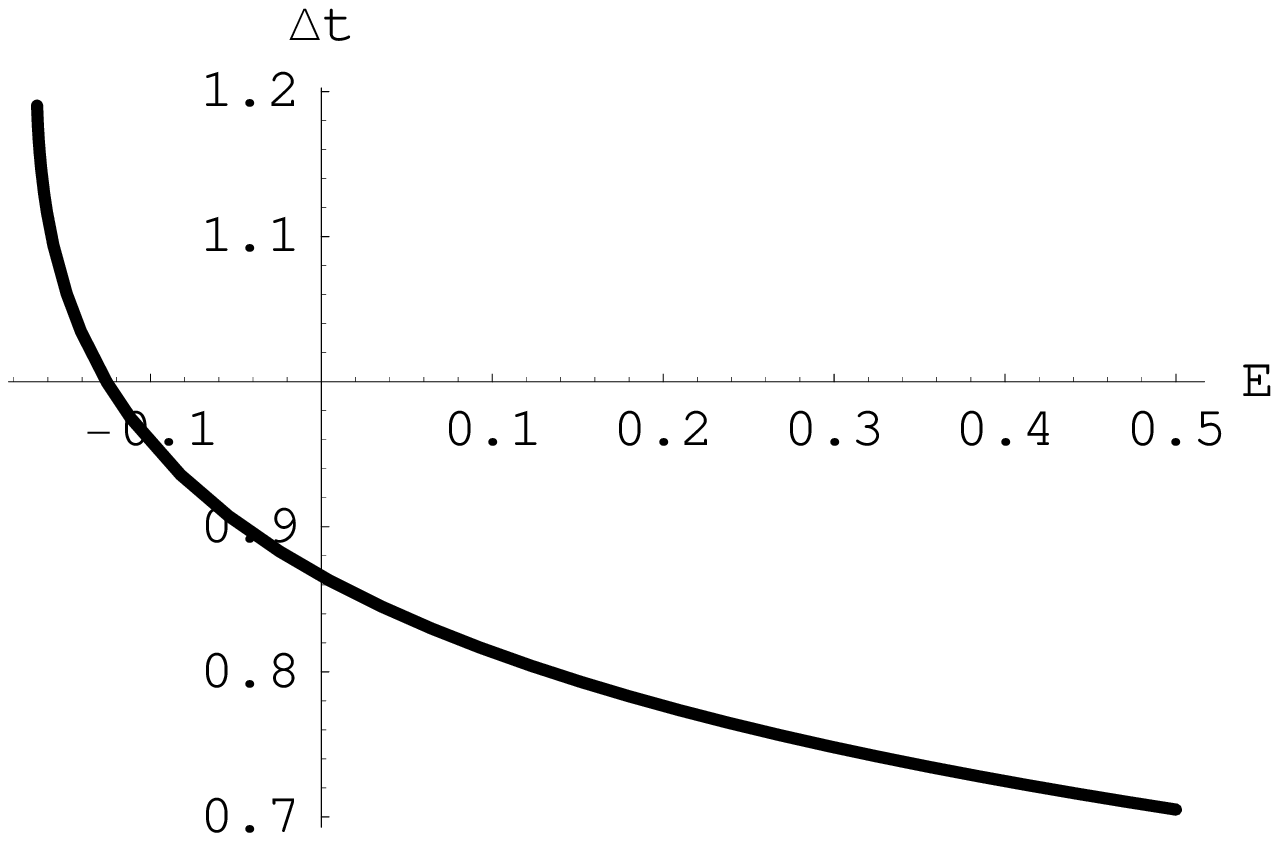,width=5
cm}{Divergence time as a function of energy for $a=1$ and $x_0=1$.
\label{Divergence_Time_Figure}}

\section{Green's function appearing in a conformal case}\label{GreenFunctiong6}
In this appendix we compute a regularized Green's function of the
operator $$L=-\Box+\al/\tau^2 ~.$$ This Green's function is used in
section \ref{ConformalCase}. For this we derive a short-distance
expansion of the Green's function which solves the equation \be
L_x\,G(x,y)=\delta(x-y) ~.\label{GreenFunctionEquation}\ee The
Green's function is given by the following expression:\be
G(x,y)=\sum\limits_n
\frac{\Psi_n(x)\Psi_n^*(y)}{\la_n},\label{GreenFunctionSum}\ee where
$\Psi$'s are eigenfunctions of $L$ and $\la$'s are corresponding
eigenvectors.

A Laplacian in $L$ is three-dimensional and involves a Euclidean
time $\tau$ and two more coordinates which we will denote
collectively by a vector $\textbf{r}$. $L$ possesses a translational
invariance in the plane of $\textbf{r}$ and therefore we are free to
put the $\delta$-function at any point in it, and we will choose
this point to be the origin. It also occurs at some time $\tau_0$.
With this choice, the Green's function will depend only on the
distance from the origin: $G(x,y)=G(\textbf{r},\tau, \tau_0)$, and
only angular-independent eigenfunctions will contribute to the sum
in (\ref{GreenFunctionSum}). So, the relevant eigenfunctions of $L$
are of the form $\Psi(x)=R(r)T(\tau)$. Here $R(r)$ satisfies the
equation \be -R''(r)-\frac 1r R'(r)=\la_r R(r) ~.\ee Normalizable
solutions of this equation exist only for positive $\la_r$ and are
given by \be R(r)=\frac{1}{\sqrt{4\pi}}J_0(\sqrt{\la_r}\,r) ~.\ee
The function $T(\tau)$ satisfies the equation \be
-T''(\tau)+\frac{\al}{\tau^2}T(\tau)=\la_{\tau} T(\tau) ~.\ee This
equation was investigated in the context of Quantum Mechanics (with
$\tau$ being a spacial coordinate) and is known to possess very
different kinds of solutions for different values of $\al$. We
consider first the case of  $\al>-1/4$. In this case $\la_{\tau}$
must be positive and the normalizable solutions are \be
T(\tau)=\sqrt{\frac{|\tau|}{2}}
\,J_{\bt}(\sqrt{\la_{\tau}}\,{|\tau|}), \qquad
\bt=\frac{\sqrt{1+4\al}}{2}.\ee In this case there are two
disconnected Hilbert spaces for different signs of $\tau$ (since, in
the language of Quantum Mechanics, there is no tunneling from a
region of $\tau>0$ to a region of $\tau<0$ and back), and therefore
the full propagator $G$ will vanish unless $\tau$ and $\tau_0$ are
of the same sign, and this is what we assume from now on.

The full normalized eigenfunction of $L$ is therefore \be
\Psi_{\la_r,\la_{\tau}}(r,\tau)=\sqrt{\frac{|\tau|}{8\pi}}\,J_{\bt}(\sqrt{\la_{\tau}}\,{|\tau|})
\,J_0(\sqrt{\la_r}\,r) ~,\ee and the corresponding eigenvalue is \be
\la=\la_r+\la_{\tau} ~.\ee

According to equation (\ref{GreenFunctionSum}), the Green's function
is given by \be
G(\textbf{r},{\tau},{\tau}_0)=\frac{\sqrt{{|\tau|}\,{|\tau|}_0}}{8\pi}
\int\limits_0^{\infty}\frac{d\la_r\,d\la_{\tau}}{\la_r+\la_{\tau}}
\,J_0({\sqrt{\la_r}\,r})\,J_{\bt}(\sqrt{\la_{\tau}}\,{|\tau|})\,J_{\bt}(\sqrt{\la_{\tau}}\,|{\tau}_0|)
~.\ee Using the integrals \cite{GradshteynRyzhik} \be
\int\limits_0^{\infty}\frac{dx}{x+y}J_0(r\sqrt{x})=2K_0(r\sqrt{y}),\ee
\be\int\limits_0^{\infty}dx\,K_0(a\,x)J_{\nu}(b\,x)J_{\nu}(c\,x)=\frac{(r_+-r_-)^{\nu}(r_++r_-)^{-\nu}}
{r_+\,r_-},\qquad r_{\pm}=\sqrt{a^2+(b\pm c)^2},\ee we end up with
the following Green's function of the operator $L$
 \be
 G(\textbf{r},{\tau},{\tau}_0)=\frac{\sqrt{{\tau}\,{\tau}_0}}{2\pi}\,\frac{1}{W_+\,W_-}\left(\frac{W_+-W_-}{W_++W_-}\right)^{\beta},\qquad
 W_{\pm}=\sqrt{\textbf{r}^2+(|{\tau}|\pm
 |{\tau}_0|)^2} ~.
 \label{GreenFunctionResult}
 \ee
This Green's function is real if $\beta$ is real, which means, for
$\al>-1/4$. If, however, $\al<-1/4$ then $\beta$ is imaginary, and
this Green's function is complex. This complex-valued Green's
function is the one that describes a rolling in the $\phi^6$
potential.

One can build a real Green's function in this case as well. In order
to do this we use a fact that there is another solution to the
Green's function equation (\ref{GreenFunctionEquation}), namely
 \be
 \tilde{G}(\textbf{r},{\tau},{\tau}_0)=\frac{\sqrt{|{\tau}|\,|{\tau}_0|}}
 {2\pi}\,\frac{1}{W_+\,W_-}\left(\frac{W_++W_-}{W_+-W_-}\right)^{\beta} ~.
 \ee
This function is not appropriate to be a Green's function in the
case of real and positive $\beta$ since it diverges for small
Euclidean times. However, for imaginary $\beta$ this function is
fine, and we can take a linear combination of it and
(\ref{GreenFunctionResult}) with equal coefficients to guarantee a
real result. The sum of coefficients must be 1 for it to indeed be a
solution to the Green's function equation. So, in the case of
imaginary $\beta$ we get the following real-valued Green's function
 \be
 \tilde{G}(\textbf{r},{\tau},{\tau}_0)=\frac{\sqrt{|{\tau}|\,|{\tau}_0}|}{2\pi}\,\frac{1}{W_+\,W_-}
 \cos\left(|\beta|\log\left(\frac{W_++W_-}{W_+-W_-}\right)\right) ~.
 \ee
This Green's function gives a zero solution in an unbounded from
below $\phi^6$ potential.

\section{Rolling in quantum mechanics}\label{QMrolling}
In this appendix we build on the Ehrenfest theorem in order to
compute the leading order quantum mechanical correction to the
Newton's equations of motion. This in turn helps one to conclude
whether quantum effects tend to accelerate or decelerate the
classical rolling.

Ehrenfest theorem reveals a quantum mechanical generalization of
Newton's second law. In particular, for a quantum particle moving in
a one-dimensional potential $V(x)$ it states
 \be
 m \frac{d^2}{dt^2}\langle \hat{x}(t) \rangle = -\langle \frac{d}{d\hat{x}}V(\hat{x}) \rangle
 \, ,
 \ee
where $m$ is the mass of the particle and $\langle \hat{x}(t)
\rangle$ is an expectation value of the position operator
$\hat{x}(t)$.

Let us define
 \bea
 x(t)&=&\langle \hat{x}(t)\rangle \, , \nonumber \\
 \hat{\eta}(t)&=&\hat{x}(t)-x(t)\, , \nonumber
 \eea
then we get
 \be
 \frac{dV}{d\hat{x}}(\hat{x}) = \frac{d V}{dx}(x(t)+\hat{\eta})
 =\frac{d V}{dx} + \frac{d^2 V}{dx^2}\hat{\eta}
 +\frac12\frac{d^3 V}{dx^3}\hat{\eta}^2\, + \, .\, .\, .
 \ee
Taking the expectation value of both sides yields
 \be
 m {d^2 \over dt^2}x(t) = -\langle {d \over d\hat{x}}V(\hat{x}) \rangle
 =-{d V \over dx}-{1 \over 2}{d^3 V \over dx^3}\sigma^2\, + \, .\, .\, ,
 \ee
where $\sigma=\sqrt{\langle \hat{\eta}^2  \rangle}$ is the standard
deviation.

Thus, in order to figure out whether quantum mechanical corrections
slow down or accelerate the rolling, one has to fix the relative
sign between the expressions on the right hand side of the last
identity. It turns out that this sign is not invariant and changes
with the shape of potential.


\section{Operator $\sB$ on radial functions}\label{SquareRootOfLaplacianInRadialCase}
In this appendix we derive a form of the operator $\sB$ when it acts
on functions which depend only on a radius and discuss some related
questions relevant to section \ref{TunnelingSection}.

If a function $\rho(x)$ depends only on the radius, then the
operator $\sB$ simplifies. Indeed, consider the Fourier transform of
$\rho(x)$. It depends only on the absolute value of the momentum $p$
and therefore $\rho(x)$ in this case can be written as \be
\rho(x)=\int\frac{d^3p}{(2\pi)^3}\,\rho(p)\,e^{-ipx}=\frac{1}{2\pi^2x}
\int\limits_0^{\infty}dp\,p\,\rho(p)\sin{p\,x} \, ,
\label{RhoExpansion}\ee where in the last equality we carried out
the angular integration. The operator $\sB$ multiplies each Fourier
mode by $|p|$, so it acts on $\rho(x)$ as
 \be
 \sB\,\rho(x)=\frac{1}{2\pi^2x}\int\limits_0^{\infty}dp\,p^2\,\rho(p)\sin{p\,x}~.
 \ee
Recall the definition of the Hilbert transform \cite{Erdelyi}
 \be
 f(x)\to
 H[f](x)=\frac{1}{\pi}PV\int\limits_{-\infty}^{\infty}\frac{f(y)}{y-x}dy ~,
 \ee
where $PV$ stands for the Cauchy principal value of the integral.
One of its main features is that it transforms $\sin{p\,x}$ to
$\textrm{sign}(p)\cos{p\,x}$ and $\cos{p\,x}$ to $-\sin{|p|\,x}$.
Using the Hilbert transform, the operator $\sB$ when it acts on a
function that depends only on a radius can be written
as\be\sB\,\rho(x)=-\frac{1}{\pi
x}PV\iii\frac{\bigl(y\rho(y)\bigr)'}{y-x}\,dy,\ee where we assumed
that the function $\rho(x)$ is defined for negative $x$ to be \be
\rho(-x)=\rho(x),\label{ParityForRho}\ee as is suggested also by
equation (\ref{RhoExpansion}).

Define, as in the body of the paper, an operator $\hA$ to be
 \be
 \hA\,\rho(x)=\frac{1}{\pi
 x}PV\iii\frac{\bigl(y\rho(y)\bigr)'}{y-x}\,dy ~.
 \ee
Find its eigenvalues and eigenfunctions. In order to do that recall
the following features of the Hilbert transform: 1) its square is
$-1$ \be H^2[f](x)\equiv H\bigl[H[f]\bigr](x)=-f(x),\ee and 2) it
commutes with a derivative \be H[f'](x)=H[f]'(x).\ee Now consider
the equation for eigenfunctions of our operator $\hA$, which can be
written as \be \hA f_k(x)\equiv\frac 1x H\bigl[(x f_k)'\bigr](x)=k
f_k(x),\label{eigenval_req}\ee where $k$ is an eigenvalue and
$f_k(x)$ is the corresponding eigenfunction. Acting on both sides
with the same linear operator once again and using the properties of
the Hilbert transform mentioned above, we arrive at the following
differential equation \be \frac
1x\bigl(x\,f_k(x)\bigr)''+k^2f_k(x)=0.\ee The solutions of this
equation are $e^{\pm i k x}/x$. Since only functions which are
finite at $x=\pm\infty$ are considered, $k$ is real. By substituting
these eigenfunctions into the equation (\ref{eigenval_req}) we find
that the corresponding eigenvalue is negative and equals to $-|k|$.
Thus one concludes that $\hat A$ is negative definite. The
eigenfunctions that possess definite parity are \be
f_k^{(1)}(x)=\frac{\sin{k x}}{x},\qquad f_k^{(2)}(x)=\frac{\cos{k
x}}{x} \, .\label{EigenFunc}\ee Since we assume that $\rho(r)$ is
even one has to drop off $f_k^{(2)}(x)$.

Next, we consider the operator $\hB$ defined similarly to equation
(\ref{TunnelingEquationWithSigma})
 \be \hB=\frac{\hA}{\arctan\,\hA} ~.\ee
This operator is positive definite, as $\hA$ is negative definite,
and can be written as \be \hB=\frac 1x \bB\,x ~,\qquad
\bB=\frac{d}{\textrm{arctanh}\, d} ~ ,\ee where $d$ means the
derivative w.r.t. $x$.

Our next goal is to both show that operator $\hB$ possesses
asymptotic eigenfunctions with negative eigenvalues, where the
definition of such a function is given in
(\ref{AsymptoticEigenfunctionDefinition}), and to present a general
procedure how to construct examples of them.

If $f(x)$ is an asymptotic eigenfunction of $\hB$ with asymptotic
eigenvalue $k$, then $g(x):=x\,f(x)$ obeys the following equation
 \be
 \lim\limits_{x\to\infty}\frac{\bB\,g(x)-k\,g(x)}{g(x)}=0 ~.
 \label{g_condition}
 \ee
Moreover, the requirement that $f(x)$ tends to zero at infinity can
be replaced by the equivalent requirement that $g(x)$ defined above
diverges at infinity slower than $x$.

Let us demonstrate how to construct the function $g(x)$ with the
above properties and negative $k$. For simplicity of notation we
take $k=-1$. By the definition of $g(x)$ one concludes that the
function $f_1(x)$ defined by
 \be
 f_1(x)=(\bB+1)g(x)\label{bB} \, , \qquad k<0
 \ee
might tend to zero at infinity faster than $g(x)$. From equation
(\ref{ParityForRho}) it follows that both functions $g(x)$ and
$f_1(x)$ have to be odd. Thus, taking an odd function $f_1(x)$ with
compact support guarantees (\ref{g_condition}) and the following
relation holds\footnote{Note that the inverse operator
$(\bB+1)^{-1}$ exists since as we have shown $\bB$ is positive
definite.}
 \be
 g(x)=(\bB+1)^{-1} f_1(x)=
 {2 \over \pi}\int\limits_0^{\infty}\Psi(k)\frac{\arctan k}{ k+\arctan k }\sin k r \,dk  ~,
 \ee
where $\Psi(k)$ is the sine-Fourier transform of $f_1(r)$
 \be
 \Psi(k)=\int_{0}^{\infty}f_1\,(r)\sin(kr)dr ~.
 \ee

\end{document}